\documentclass[fleqn,usenatbib]{mnras}

\usepackage{newtxtext,newtxmath}

\usepackage[T1]{fontenc}
\usepackage{ae,aecompl}

\usepackage{graphicx}	
\usepackage{amsmath}	
\usepackage{siunitx}    
\usepackage{lipsum}

\DeclareSIUnit \jy {Jy}
\DeclareSIUnit \pc {pc}
\DeclareSIUnit \htwo {H\mbox{$_2$}}
\DeclareSIUnit \msol {M\mbox{$_\odot$}}
\DeclareSIUnit \Av {A\mbox{$_\mathrm{v}$}}
\DeclareSIUnit \au {AU}
\newcommand{\Herschel}{\textit{Herschel}}

\newcommand{\AoI}{S1-HD147889 axis}

\title[Filaments in Ophiuchus]{A PPMAP analysis of the filamentary structures in Ophiuchus L1688 and L1689}

\author[A. D. P. Howard, et al.]{A. D. P. Howard,$^{1}$\thanks{E-mail: Alexander.Howard@astro.cf.ac.uk (CU)}
A. P. Whitworth,$^{1}$
M. J. Griffin,$^{1}$
K. A. Marsh$^{2}$
and M. W. L. Smith$^{1}$
\\
$^{1}$School of Physics and Astronomy, Cardiff University, 5 The Parade, Cardiff, CF24 3AA, UK\\
$^{2}$IPAC, Caltech, 1200E California Boulevard, Pasadena, CA 91125, USA\\
}

\date{Accepted XXX. Received YYY; in original form ZZZ}

\pubyear{2020}

\begin{document}
\label{firstpage}
\pagerange{\pageref{firstpage}--\pageref{lastpage}}
\maketitle

\begin{abstract}
We use the PPMAP (Point Process MAPping) algorithm to re-analyse the \textit{Herschel} and SCUBA-2 observations of the L1688 and L1689 sub-regions of the Ophiuchus molecular cloud. PPMAP delivers maps with high resolution (here $14''$, corresponding to $\sim 0.01\,{\rm pc}$ at $\sim 140\,{\rm pc}$), by using the observations at their native resolutions. PPMAP also delivers more accurate dust optical depths, by distinguishing dust of different types and at different temperatures. The filaments and prestellar cores almost all lie in regions with $N_{\rm H_2}\gtrsim 7\times 10^{21}\,{\rm cm}^{-2}$ (corresponding to $A_{_{\rm V}}\gtrsim 7$). The dust temperature, $T$, tends to be correlated with the dust opacity index, $\beta$, with low $T$ and low $\beta$ tend concentrated in the interiors of filaments. The one exception to this tendency is a section of filament in L1688 that falls -- in projection -- between the two B stars, S1 and HD147889; here $T$ and $\beta$ are relatively high, and there is compelling evidence that feedback from these two stars has heated and compressed the filament. Filament {\sc fwhm}s are typically in the range $0.10\,{\rm pc}$ to $0.15\,{\rm pc}$. Most filaments have line densities in the range $25\,{\rm M_{_\odot}\,pc^{-1}}$ to $65\,{\rm M_{_\odot}\,pc^{-1}}$. If their only support is thermal gas pressure, and the gas is at the canonical temperature of $10\,{\rm K}$, the filaments are highly supercritical. However, there is some evidence from ammonia observations that the gas is significantly warmer than this, and we cannot rule out the possibility of additional support from turbulence and/or magnetic fields. On the basis of their spatial distribution, we argue that most of the starless cores are likely to disperse (rather than evolving to become prestellar).
\end{abstract}


\section{Introduction}\label{sec:intro}

In the past decade, far-infrared and sub-millimetre dust observations have greatly enhanced our understanding of the star formation process within the Milky Way. Extensive surveys conducted with telescopes such as \textit{Herschel}\footnote{\textit{Herschel} is an ESA  space observatory with science instruments provided by European-led Principal Investigator consortia and with important participation from NASA.} and the JCMT have helped to disentangle the complicated nature of molecular clouds, revealing widespread networks of dense filaments \citep{Andre2010, Arzoumanian2011, Hacar2013, Konyves2015, Arzoumanian2019, Ladjelate2020}, and linking them to the earliest stages of pre- and protostellar core formation \citep{Andre2014, Pattle2015, Marsh2016, Ladjelate2020}.

The Ophiuchus molecular cloud complex is a nearby \citep[\SI{\sim 140}{\pc},][]{Mamajek2008}, well studied region of star formation associated with the Gould Belt \citep{Wilking1983, Nutter2006, Pattle2015, Soler2019, Ladjelate2020}. The complex can be broken into a number of visually distinct sub-regions, the two most massive of which are L1688 and L1689. The Sco OB2 association, which is just $11 \pm 3\,{\rm pc}$ away \citep{Mamajek2008}, is thought to have a strong influence on the region, driving elongated streamers away from the dense heart of the complex \citep{Vrba1977, Loren1989}. The L1688 sub-region, which is closer to Sco OB2, appears to be forming stars more actively than L1689 \citep{Nutter2006}.

Despite the early identification of filamentary structures associated with the Ophiuchus complex, most studies have focused on the pre- and protostellar cores within L1688 and L1689 \citep[e.g.][]{Nutter2006, Pattle2015}, due to the limitations associated with observing extended structures with ground-based telescopes. Recent studies have analysed the filament networks, but have tended to do so in general terms, presenting the average properties of the entire ensemble of structures \citep[e.g.][]{Arzoumanian2019, Ladjelate2020}.

In this paper, we re-analyse \textit{Herschel} and SCUBA-2 observations of the L1688 and L1689 sub-regions of the Ophiuchus cloud complex, using the Bayesian PPMAP algorithm \citep{Marsh2015}. Section \ref{sec:obs} describes the observations of the Ophiuchus complex. Section \ref{sec:approx} lists the approximations that are made, and derives the factor for converting dust optical depths into surface densities (in $\rm{M_{_\odot}\,pc^{-2}}$) and column-densities of molecular hydrogen (in $\rm{cm^{-2}}$). Section \ref{sec:ppmap-methods} outlines the operation of the PPMAP algorithm. Section \ref{sec:ppmap-prod} presents the basic PPMAP data products for L1688 and L1689, and discusses the wide scale variations in dust properties across the two sub-regions. Section \ref{sec:mass} investigates and compares the mass distributions and core formation efficiencies in L1688 and L1689. Section \ref{sec:filaments} gives a detailed analysis of the network of filaments across Ophiuchus, and examines the impact of feedback from nearby B stars on the largest filamentary structure. Section \ref{sec:conc} summarises our main conclusions.

\section{Observations of Ophiuchus}\label{sec:obs}

\subsection{Herschel observations}\label{sec:obs:herschel}

The \textit{Herschel} observations of the Ophiuchus molecular cloud were taken as part of the {\em Herschel Gould Belt Survey} (HGBS)\footnote{\url{http://www.herschel.fr/cea/gouldbelt/en/}}. They consist of {$70\mu{\rm m}$ and $160\mu{\rm m}$} data from PACS \citep{Poglitsch2010}, and {$250\mu{\rm m}$, $350\mu{\rm m}$ and $500\mu{\rm m}$} data from SPIRE \citep{Griffin2010}. The scans were taken in the fast scan ($60''\,\rm{s}^{-1}$) PACS/SPIRE parallel mode, and comprise a single pair of nominal and orthogonal scans, both taken on 25 September 2010.  The {\em Herschel} Observation IDs for these scans are 1342205093 and 1342205094, respectively. 

The observations were reduced using the HIPE User Release v15.0.1. PACS maps were produced using a modified version of the \verb|JScanam| task, whilst SPIRE maps were produced with the \verb|mosaic| script operating on Level 2 data products. We determined the Zero-Point Offsets for the PACS maps through comparison with \textit{Planck} and \textsc{iras} observations of the same region \citep[cf.][]{Bernard2010}. The adopted offsets were $-8.4\,\rm{MJy\,sr^{-1}}$ and $243.5\,\rm{MJy\,sr^{-1}}$ for the \SIlist{70;160}{\micro \meter} bands, respectively.\footnote{These are different from the offsets used by \citet{Ladjelate2020} (their Table 1), possibly because of the different smoothing they apply and/or differences in the reduction procedure used on the PACS data. If we were to adopt the \citet{Ladjelate2020} offsets, we would obtain lower dust temperatures and therefore larger dust optical depths. However, this would be partially compensated by the fact that PPMAP does not overestimate the contribution from warmer than average dust.} The SPIRE Zero-Point Offsets were applied automatically as part of the reduction process.

The {\sc fwhm}s of the circular beam profiles used by PPMAP are \SI{8.5}{\arcsecond}, \SI{13.5}{\arcsecond}, \SI{18.2}{\arcsecond}, \SI{24.9}{\arcsecond}, and \SI{36.3}{\arcsecond}, for -- respectively -- the \SI{70}{\micro \meter}, \SI{160}{\micro \meter}, \SI{250}{\micro \meter}, \SI{350}{\micro \meter}, and \SI{500}{\micro \meter} wavebands \citep{Exter2017,Valtchanov2017}. In reality, the PACS beams are distorted by the fast scan PACS/SPIRE parallel mode, producing effective non-circular beamsizes of $\sim 6\si{\arcsecond} \times 12\si{\arcsecond}$ for the \SI{70}{\micro \meter} waveband, and $\sim 12\si{\arcsecond} \times 16\si{\arcsecond}$ for the \SI{160}{\micro \meter} waveband, but this is not taken into account in the standard PPMAP procedure.

\subsection{SCUBA-2 observations}\label{sec:obs:scuba2}

As in \citet{Howard2019}, we use SCUBA-2 \citep{Holland2013} \SI{850}{\micro \meter} observations to supplement the \textit{Herschel} data. These observations were taken as part of the JCMT Gould Belt Survey \citep{Ward-Thompson2007}, and have an angular resolution of \SI{14.6}{\arcsecond}. The observations consist of \SI{30}{\arcminute} diameter circular regions generated using the PONG1800 mapping mode \citep{Chapin2013}, and were mosaicked together as described in \citet{Pattle2015}.

A \SI{10}{\arcminute} high-pass filter was used in the reduction of the SCUBA-2 observations to remove the effects  of atmospheric and instrumental noise. This has the effect of removing emission from structure larger than \SI{10}{\arcminute} in extent. We restore the larger spatial scales by combining the SCUBA-2 observations with the \SI{850}{\micro \meter} \textit{Planck} map of the region, using a customised feathering technique in \verb|python|. The feathering uses the same core routine as the \textsc{CASA feather} task\footnote{https://casa.nrao.edu/casadocs/casa-5.4.1/image-combination/feather\#cit1}, where both maps are first converted to the Fourier-plane, and then the low-resolution image is scaled by the ratio of the beam volumes. The Fourier-transformed high-resolution image is then weighted by $\left(1 - {\rm FT}_{\rm beam}\right)$ where ${\rm FT}_{\rm beam}$ is the Fourier-transform of the low-resolution image beam, and in this code we assume the beams are Gaussian. Finally the high and low-resolution datasets are added and inverse Fourier-transformed to produce the combined image. Tests of this script on extra-galactic targets show that the recovered flux-densities are accurate to $\sim$10\%. Moreover, the maps presented here are likely to be better than this, since the SCUBA-2 maps preserve emission scales up to 10$^\prime$ scales, whereas extragalactic maps typically use a harsher filter ($\sim6^\prime$). The algorithm, and tests of its fidelity when applied to SCUBA-2 data, will be submitted for publication shortly (Smith et al., in prep.).

\begin{figure*}
    \centering
    \includegraphics[width = \textwidth]{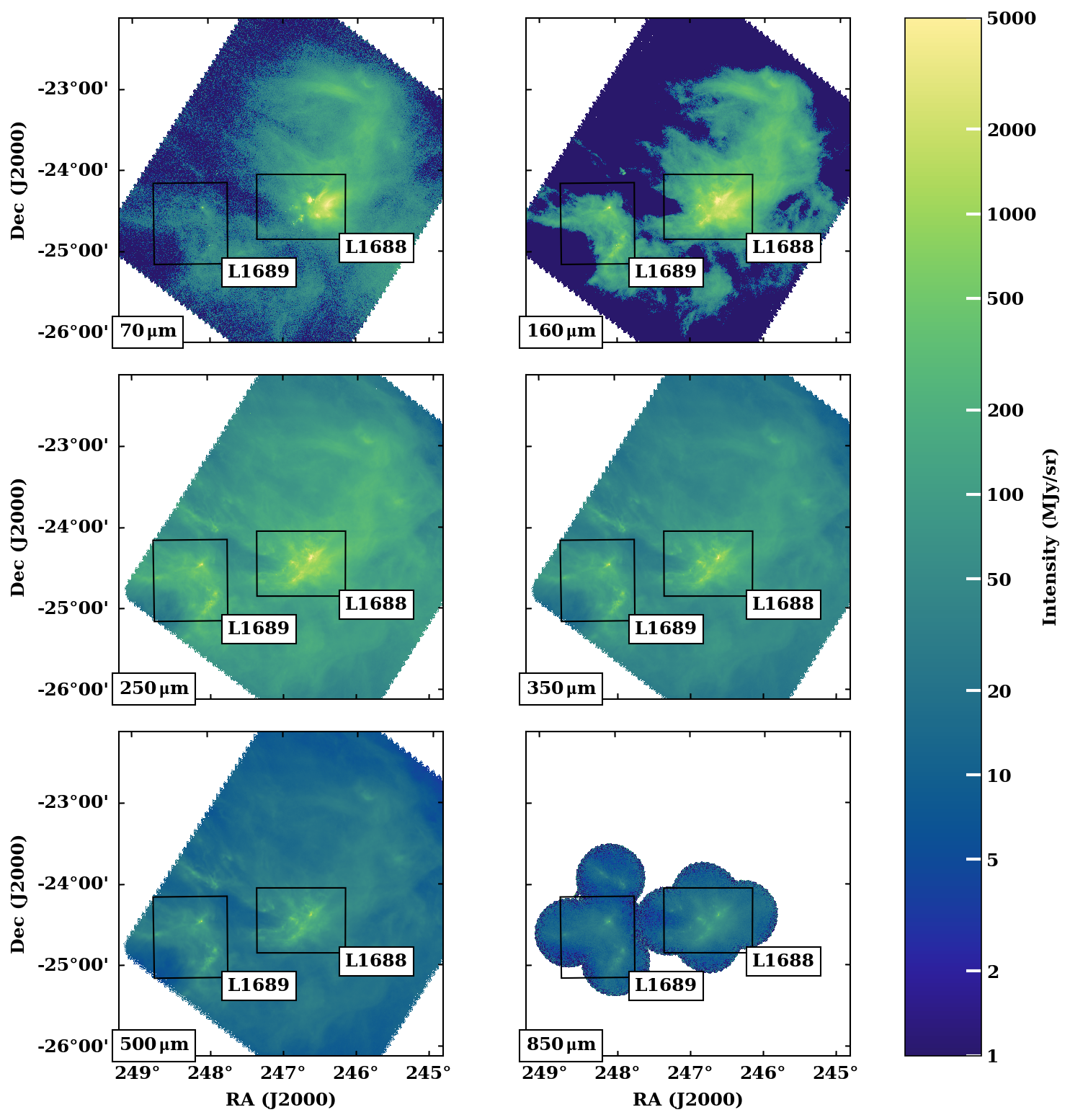}
    \caption{Intensty maps of Ophiuchus in the {\it Herschel} bands at \SI{70}{\micro \metre}, \SI{160}{\micro \metre}, \SI{250}{\micro \metre}, \SI{350}{\micro \metre} and \SI{500}{\micro \metre}, plus the SCUBA-2 band at \SI{850}{\micro \metre}. The black rectangles delineate the L1688 and L1689 sub-regions.}
    \label{fig:obsChart}
\end{figure*}

Fig.~\ref{fig:obsChart} shows the \Herschel{} and SCUBA-2 observations used in this paper, and the boundaries of the sub-regions encompassing the L1688 and L1689 objects \citep[see][]{Lynds1962}. As the boundaries of L1688 and L1689 are not well defined, we chose the coverage of these sub-regions based on the extent of the SCUBA-2 observations, seeking to exclude areas that have not been observed, while at the same time enclosing the brightest regions. The L1688 sub-region has an angular size of \SI{1.2}{\degree} by \SI{0.8}{\degree} centred on RA \SI{=246.75}{\degree} and Dec \SI{=-24.47}{\degree}. The L1689 region encompasses an area of \SI{1.0}{\degree} by \SI{1.0}{\degree} centred on RA \SI{=248.25}{\degree} and Dec \SI{=-24.67}{\degree}.

\section{Approximations}\label{sec:approx}

\subsection{Opacity Law}

We follow the convention of approximating the long-wavelength variation of the mass opacity, $\kappa_\lambda$, with wavelength, $\lambda$, as a power law with opacity index, $\beta$:
\begin{equation}\label{eq:opacityLaw}
\kappa_\lambda=\kappa_0 \left( \frac{\lambda}{\lambda_0} \right)^{-\beta}.
\end{equation}
Here $\kappa_0$ is the mass opacity (per unit mass of dust and gas) at the reference wavelength, for which we adopt $\lambda_0=300\,\mu\rm{m}$.

\subsection{Conversion factors}\label{SEC:ConvFact}

The optical depth, $\tau_0$, at the reference wavelength, $\lambda_0$, is related to the surface-density of gas and dust, $\Sigma$, by $\tau_0 = \Sigma \kappa_0$, and hence
\begin{equation}\label{EQN:Sigma.1}
\Sigma=\frac{\tau_0}{\kappa_0}\,.
\end{equation}
If the fractional abundance of hydrogen by mass is $X_{\rm H}$, and the fraction of hydrogen that is molecular is $X_{\rm H_2}$, the column-density of molecular hydrogen is
\begin{equation}\label{EQN:NH2.1}
N_{\si{\htwo}} = \frac{X_{\rm H} X_{\rm H_2} \Sigma}{2 m_\mathrm{H}} = \frac{X_{\rm H} X_{\rm H_2} \tau_0}{2 m_\mathrm{H} \kappa_0},
\end{equation}
where $m_\mathrm{H}=1.67\times 10^{-24}\,\rm{g}$ is the mass of an hydrogen atom.

However, the fundamental quantity estimated by PPMAP from the analysis of \Herschel{} and SCUBA-2 maps is the dust optical-depth, $\tau_0$, at the reference wavelength, $\lambda_0$. Converting $\tau_0$ into $\Sigma$ (Eq. \ref{EQN:Sigma.1}) or $N_\mathrm{H_2}$ (Eq. \ref{EQN:NH2.1}) requires the specification of $\kappa_0$,  $X_{\rm H}$ and $X_{\rm H_2}$, all of which are uncertain and all of which are expected to vary with position -- especially $\kappa_0$ and $X_{\rm H_2}$. Parenthetically these parameters also have no bearing on the workings of {\sc ppmap}. For those who wish to convert $\tau_0$ into $\Sigma$, we suggest $\kappa_0=0.10\,{\rm cm^2\,g^{-1}}=2.1\times 10^{-5}\,{\rm pc^2\,M_\odot^{-1}}$, which is consistent with the opacity coefficient proposed by \citet{Hildebrand1983} for a fractional dust abundance of $Z_\mathrm{D} = 0.01$, and gives
\begin{equation}\label{EQN:tau2Sigma}
\Sigma=\left[4.8\times 10^4\,{\rm M_\odot\,pc^{-2}}\right]\;\tau_0\,.
\end{equation}
For those who wish to convert $\tau_0$ into $N_\mathrm{H_2}$, we suggest $X_{\rm H}=0.70$ and $X_{\rm H_2}\simeq1.00$, whence
\begin{equation}\label{EQN:tau2NH2}
N_{\rm H_2}=\left[2.1\times 10^{24}\,{\rm cm^{-2}}\right]\;\tau_0\,.
\end{equation}

The total mass on the line of sight through a pixel of angular size $\Delta\Omega_{\rm pixel}$ is 
\begin{equation}
\begin{split}
\Delta M_{\rm pixel}&=\Sigma\,(\Delta\Omega_{\rm pixel}/{\rm steradian})\,D^2\\\label{EQN:massFromCdens}
&=\left[0.011\,{\rm M}_\odot\right]\;\tau_0\;\left(\!\frac{\Delta\Omega_{\rm pixel}}{{\rm arcsec}^2}\!\right)\;\left(\!\frac{D}{100\,\rm{pc}}\!\right)^{\!2}\!,
\end{split}
\hspace{0.5cm}
\end{equation}
where $D$ is the distance to the observed region and $\tau_0$ is the optical depth through the pixel, at the reference wavelength.

\subsection{Caveats}

The factors in square brackets in Eqs. (\ref{EQN:tau2Sigma}) through (\ref{EQN:massFromCdens}) are \textit{not} accurate to two significant figures. Furthermore, when we derive variations in the opacity index, $\beta$, we should be mindful that these variations are probably due to grain growth/erosion and/or coagulation/fragmentation. Such changes are likely to be accompanied by correlated changes in the dust absorption opacity at the reference wavelength, $\kappa_0$. The magnitudes of these changes are not currently known, and even their sense is not established with total certainty. This uncertainty does not affect the variations in $\beta$ which we detect, but it does affect the amount of mass ($\Sigma$) or molecular hydrogen ($N_{\si{\htwo}}$) associated with the different types of dust. Thus, when we refer to the line-of-sight mean dust opacity index, $\bar{\beta}$ (e.g. Eq. \ref{EQN:beta_bar}), or the line-of-sight mean temperature, $\bar{T}$ (e.g. Eq. \ref{EQN:T_bar}), we should be mindful that these are optical-depth weighted means, and not -- strictly speaking -- mass-weighted means. In contrast, the means returned by conventional modified black body fitting routines are flux-weighted means.

\section{The PPMAP Algorithm}\label{sec:ppmap-methods}

In the standard Modified BlackBody (MBB) fitting procedure the observations in the different wavebands are smoothed to the coarsest resolution (here the $36.3''$ of the {\it Herschel} \SI{500}{\micro \meter} waveband), thereby discarding a large amount of information. The intensity in each pixel is then {fitted} with a single average optical depth, ${\bar\tau}_0$, and a single average dust temperature, ${\bar T}$ (i.e. flux-weighted averages along the line of sight through the pixel),
\begin{equation}\label{EQN:MBBInt}
I_\lambda={\bar{\tau}_0}\left(\!\frac{\lambda}{\lambda_0}\!\right)^{\!-\beta}B_{\lambda}\!\left({\bar T}\right)\,,
\end{equation}
with $\beta\!=\!2$ \citep[e.g.][]{Hildebrand1983, Andre2010, Schneider2012, Konyves2015, Ladjelate2020}. In principle, the emissivity index, $\beta$, can also {be} treated as a free parameter, but in practice this endeavour is compromised by the fact that lower values of $\beta$ are very hard to distinguish from higher values of $T$, and vice versa \citep{Shetty2009a, Shetty2009b}, particularly if there are no observations at long wavelengths well above the peak of the SED. The resolution can be improved by a factor of two by bootstrapping off the {\it Herschel} \SI{250}{\micro \meter} observations \citep{Palmeirim2013}, but this is only reliable if there is a low temperature variance along the line of sight. The maps of Ophiuchus presented in Ladjelate with \SI{18}{\arcsecond} resoloution are obtained using the Palmeirim bootstrapping procedure

In contrast to standard MBB fitting procedure, PPMAP \citep[PointProcessMAPping;][]{Marsh2015} does not degrade the set of input observations to the coarsest common resolution. Instead, it utilises high-fidelity beam profiles for each of the observing bands, and thereby retains some of the extra information contained within the observations at their native resolutions.

The PPMAP data products derived here for the Ophiuchus sub-regions have \SI{14}{\arcsecond} angular resolution in order to make the best use of the information provided by the PACS \SI{70}{\micro \meter} observations (with a distorted minimum resolution of \SI{12}{\arcsecond}) and the SCUBA-2 \SI{850}{\micro \meter} observations (with a resolution of \SI{14.6}{\arcsecond}). The average distance to the Ophiuchus sub-regions is $D \simeq 140$\,\si{\pc} \citep{Mamajek2008}. Therefore, the \SI{14}{\arcsecond} resolution corresponds to $\sim$\,\SI{0.01}{\pc} (or $\sim$\,\SI{2000}{AU}). In principle, PPMAP can deliver maps at the finest observed resolution (i.e. the \SI{8.5}{\arcsecond} of the \SI{70}{\micro \meter} waveband, as demonstrated in the analysis of M31 presented in \citet{Whitworth2019}), but this requires higher signal-to-noise than we have for Ophiuchus.

PPMAP presumes that the total emission observed in each pixel is likely to involve contributions from different populations of dust along the line of sight, with different absorption properties and different dust temperatures. The differing dust absorption properties are represented by different discrete values of $\beta$, labelled $\beta_{k}$. In this work we use three linearly spaced values, $\beta_{1}\!=\!1.0,\;\beta_{2}\!=\!1.5$ and $\beta_3\!=\!2.0$.\footnote{In exploratory work we invoked a fourth discrete $\beta$ value, $\beta_4=2.5$, but the contribution to the optical depths from this dust was negligible, and therefore we dropped it from consideration.} While each value of $\beta_k$ is a discrete, delta function, the values are intended to encompass all opacity indices within a small interval centred on $\beta_k$; for example, $\beta_1 = 1.0$ represents the opacity index range from $0.75$ to $1.25$.

Similarly, the different dust temperatures are represented by a series discrete values, labelled $T_{\ell}$. Here we use twelve logarithmically spaced values, 
$T_1\!=\!7.0\,\si{\kelvin},\,
T_2\!=\!8.4\,\si{\kelvin},\,
T_3\!=\!10.0\,\si{\kelvin},\,
T_4\!=\!12.0\,\si{\kelvin},\,
T_5\!=\!14.3\,\si{\kelvin},\,
T_6\!=\!17.1\,\si{\kelvin},\,
T_7\!=\!20.5\,\si{\kelvin},\,
T_8\!=\!24.5\,\si{\kelvin},\,
T_9\!=\!29.2\,\si{\kelvin},\,
T_{10}\!=\!35.0\,\si{\kelvin},\,
T_{11}\!=\!41.8\,\si{\kelvin},$ and $T_{12}\!=\!50.0\,\si{\kelvin}$.
Again, $T_{\ell}$ is intended to represent a small range of dust temperature; for example, $T_1\!=\!7.0\,\si{\kelvin}$ represents the dust temperature range from \SIrange{6.4}{7.7}{\kelvin}, and similarly for the other $T_{\ell}$ values.

The PPMAP algorithm assumes that the dust emission is optically thin. Therefore optically thick regions, such as protostellar cores \citep{Ossenkopf1994}, are poorly fit by the algorithm. In cases where the emission from such cores dominates over the cloud emission, they must be masked out before the observations are analysed. Initial tests have demonstrated that in Ophiuchus, masking is only necessary for the L1689-IRS6 Class I protostar \citep{Greene1994} and the 16293-2422 binary/multiple Class 0 protostellar system \citep{Mundy1992}. A \SI{40}{\arcsecond} diameter circular mask has been applied to each of the observation bands at the centroid positions of these two sources. The masks appear as small white circles on the PPMAP maps of the L1689 sub-region. There are other protostellar cores that are optically thick (for example at ${\rm RA}=246^\circ~48,\;{\rm Dec}=-24^\circ\,19.5'$ on Fig. 4), but their effect is sufficiently small and local that we can reasonably ignore them.

PPMAP assumes that the observed intensity in each pixel on the sky, $(i,j)$, can be approximated by
\begin{equation}\label{EQN:PPInt}
I_{\lambda}=
\sum\limits_{k=1}^{k={3}}\,
\sum\limits_{\ell=1}^{\ell=12}\,
\left\{\Delta^2\tau_{{0:k\ell}}\;
\left(\!\frac{\lambda}{\lambda_0}\!\right)^{\!-\beta_{k}}
\;B_{\lambda}\!\left(T_{\ell}\right)\right\},
\end{equation}
where $\Delta^2\tau_{{0:k\ell}}$ is the contribution to the total optical depth, $\tau_0$, at the reference wavelength,  $\lambda_0 = 300\,\si{\micro \meter}$, from dust along the line of sight with $\beta\sim\beta_{k}$ and $T\sim T_{\ell}$. 

PPMAP constructs a model of the distribution of dust optical depth based on these assumptions. Therefore the raw data products from PPMAP are four-dimensional data-hypercubes, with two dimensions representing position on the sky, $(x_i,y_j)$, one dimension representing the opacity index, $\beta_{k}$, and one dimension representing the dust temperature, $T_{\ell}$. Two data-hypercubes are produced, one giving the expectation values for $\Delta^2\tau_{{0:k\ell}}$, and the other giving the corresponding uncertainties $\Delta^2\sigma_{{0:k\ell}}$.

The data-hypercubes are produced using a Bayesian point process algorithm. Initially, the algorithm populates the optical depth data-hypercube with a uniform array of very small optical-depth quanta, $\delta\tau_{300\,\si{\micro \meter}}$. The emission that such a distribution would produce in each of the observation wavebands is compared with the true observations, taking into account the effects of the instrument beam profiles and colour corrections, and initially assuming an extremely high -- artificially inflated -- level of noise. The distribution of optical-depth quanta is then adjusted to improve the fit between the predicted intensities and the observed intensities. Due to the initially high level of artificial noise, the fit seems quite good and so the adjustments are small, i.e. in the linear regime. The comparison and adjustment process is performed iteratively, and the artificial noise is reduced with each iteration until it is completely removed and the predicted intensities (i.e. Eq. \ref{EQN:PPInt}) closely match the observed intensities. Details of the algorithm are given in \citet{Marsh2015}, along with a range of tests on synthetic data. The algorithm invokes a tight, Gaussian prior on $\beta$, in order to mitigate the $(\beta,T)$ degeneracy \citep[i.e. the fact that, for data sets with limited wavelength range, low $\beta$ can be mimicked by high $T$, and vice versa;][]{Shetty2009a,Shetty2009b}.  For Ophiuchus, we adopt a prior with mean $\mu_{\beta}\!=\!2.0$ and standard deviation $\sigma_{{\!\beta}}\!=\!0.25$. The tight prior on $\beta$ ensures that the algorithm only deviates from the canonical value of $\beta\!=\!2.0$ when the data really require this. A flat prior is used for $\log(T)$.

\begin{figure*}
    \centering
    \includegraphics[width = 1.0\textwidth]{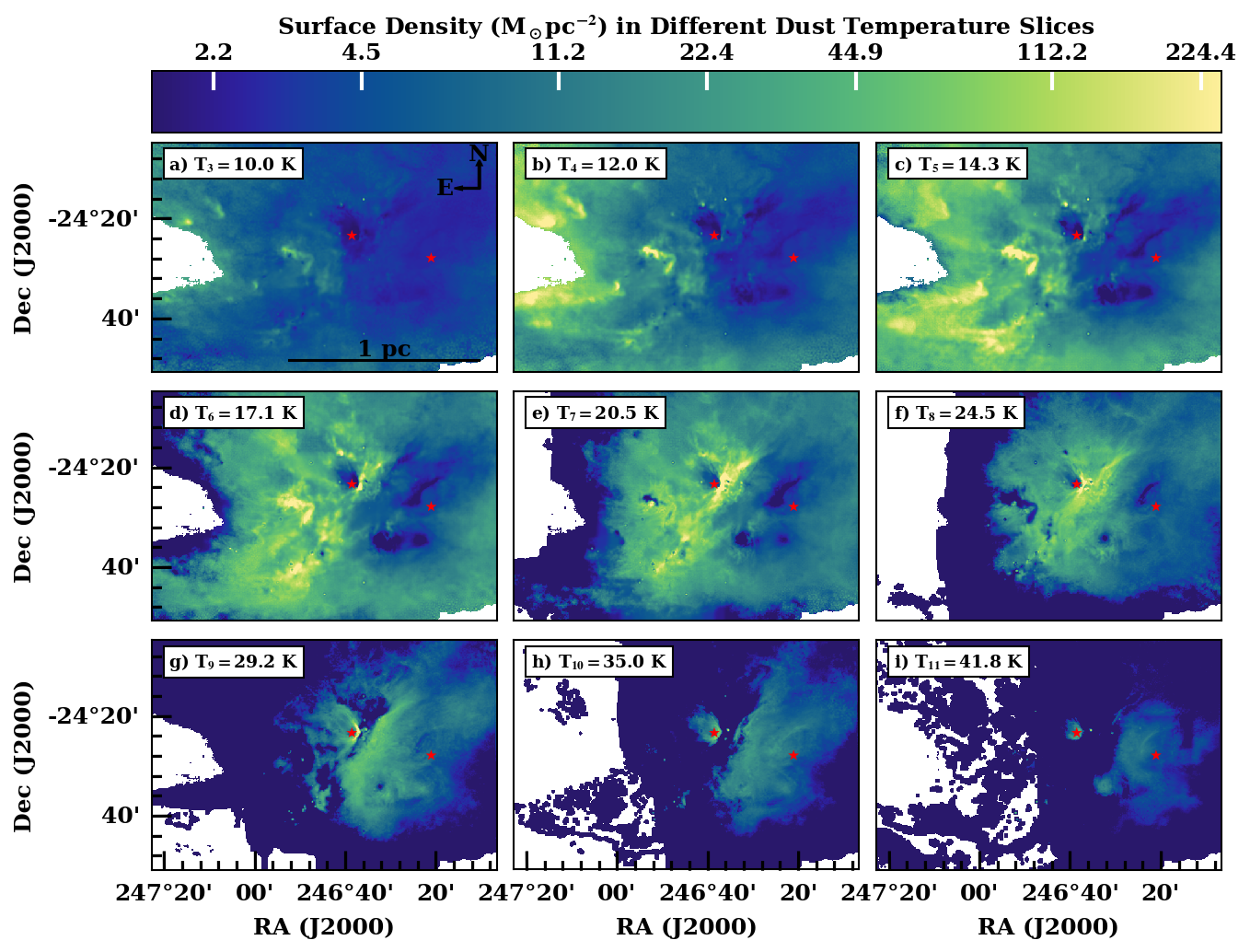}
    \caption{Nine contiguous PPMAP temperature slices for the L1688 sub-region, showing the estimated surface density, $\Sigma$, as traced by dust having temperature represented by the value marked in the top left. The red stars indicate the positions of the S1 and HD147889 pre-main sequence B stars.}
    \label{fig:L1688TempGrid}
\end{figure*}

\begin{figure*}
    \centering
    \includegraphics[width = 1.0\textwidth]{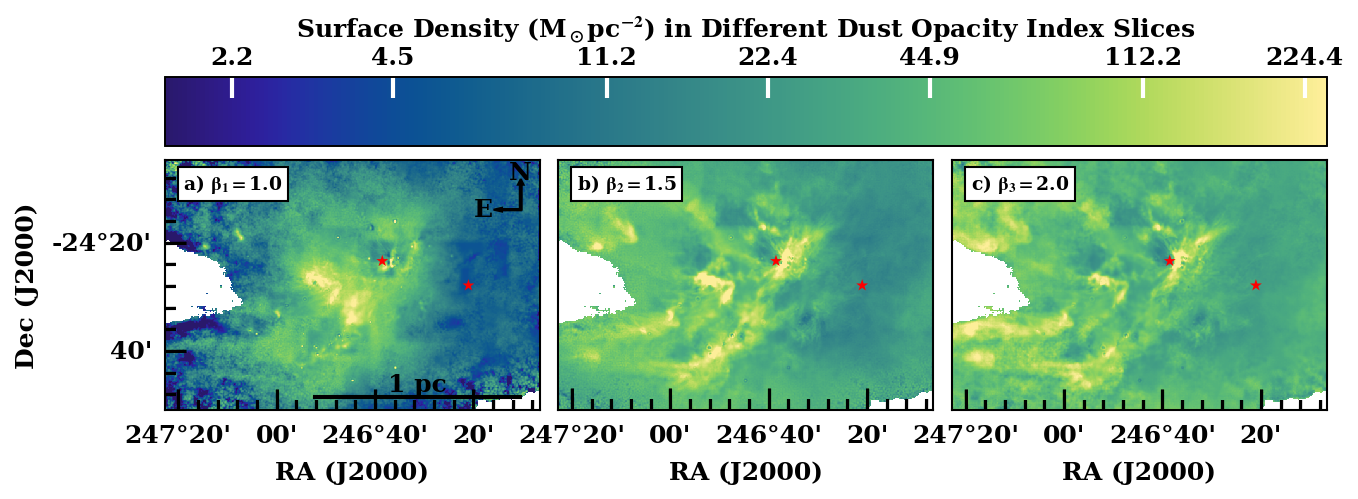}
    \caption{Three contiguous PPMAP opacity-index slices for the L1688 sub-region, showing the estimated surface density, $\Sigma$, as traced by dust with $\beta$ close to the value marked in the top left. The red stars indicate the positions of the S1 and HD147889 pre-main sequence B stars.}
    \label{fig:L1688BetaGrid}
\end{figure*}

\begin{figure*}
    \centering
    \includegraphics[width = 1.0\textwidth]{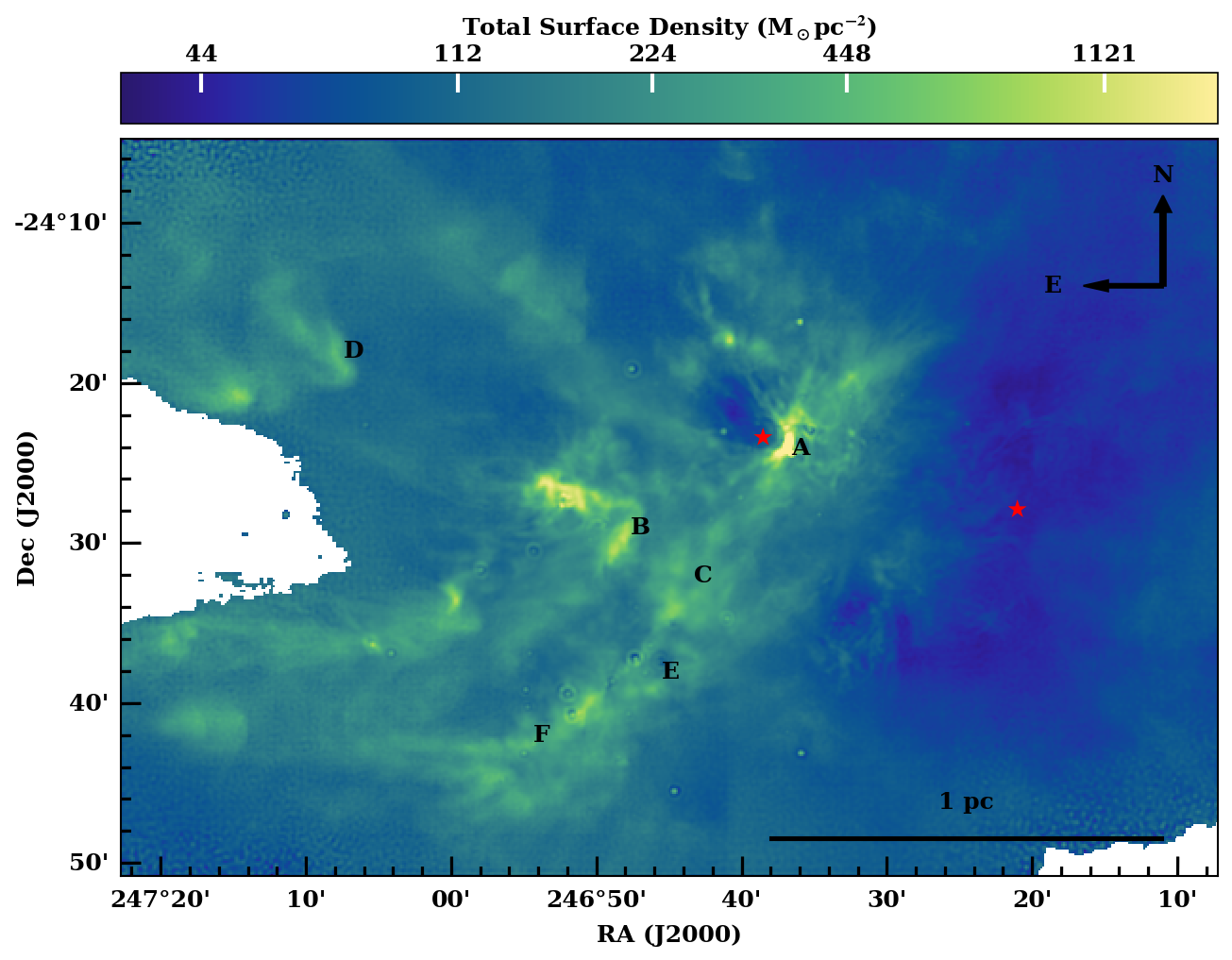}
    \caption{The estimated total surface density, $\Sigma$, for the L1688 sub-region. Red stars indicate the positions of the S1 and HD147889 pre-main sequence B stars. Letters A-F indicate the approximate positions of the dense DCO$^+$ cores identified by \citet{Loren1990}.}
    \label{fig:L1688Cdens}
\end{figure*}

\begin{figure*}
    \centering
    \includegraphics[width = 1.0\textwidth]{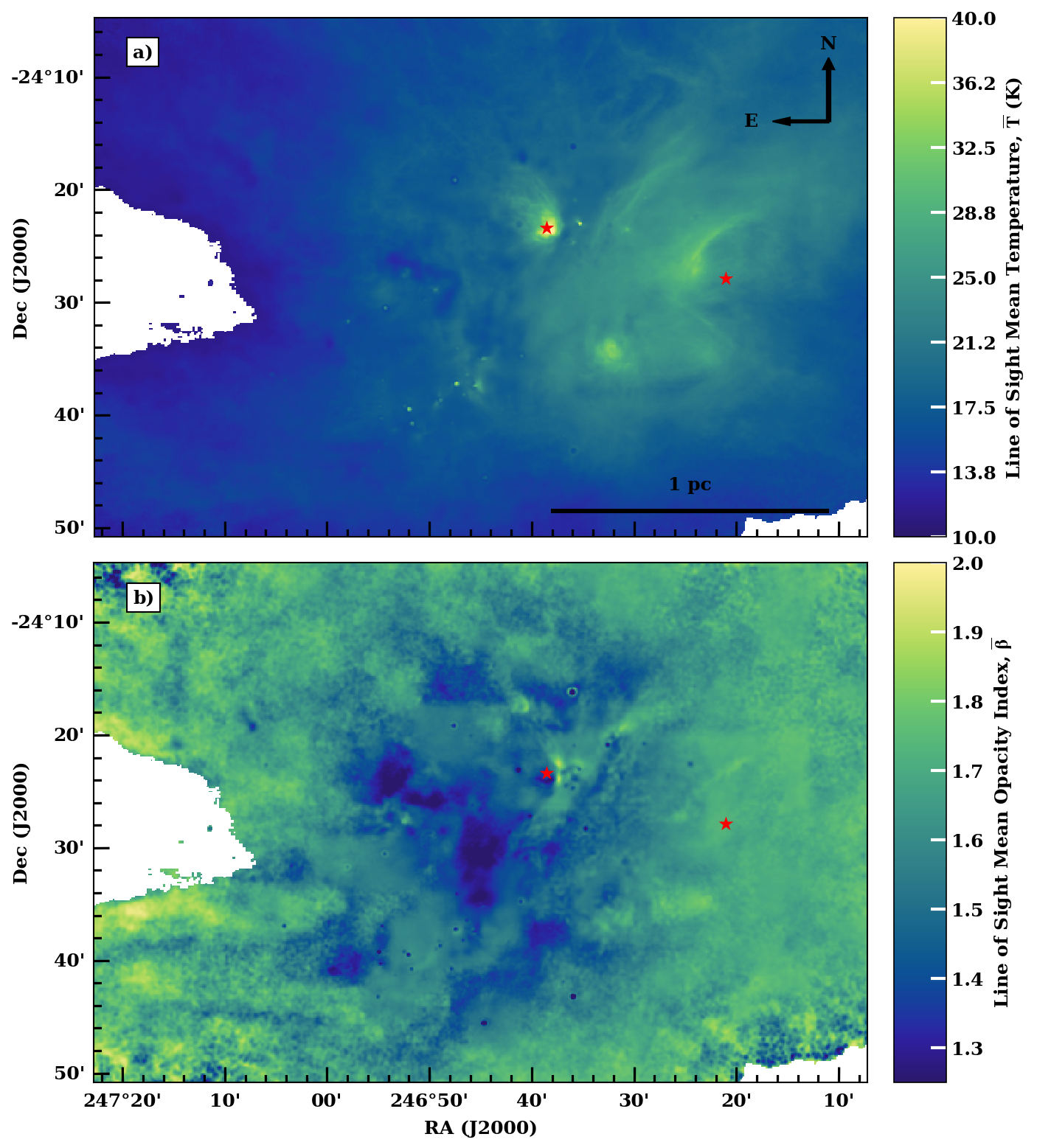}
    \caption{Maps of (a) the mean line-of-sight dust temperature (Eq. \ref{EQN:T_bar}), and (b) the mean line-of-sight dust opacity index (Eq. \ref{EQN:beta_bar}), for the L1688 sub-region. The red stars indicate the positions of the S1 and HD147889 pre-main sequence B stars.}
    \label{fig:L1688TB}
\end{figure*}

\begin{figure*}
    \hspace{-1.0cm}\includegraphics[height = 0.80\textheight]{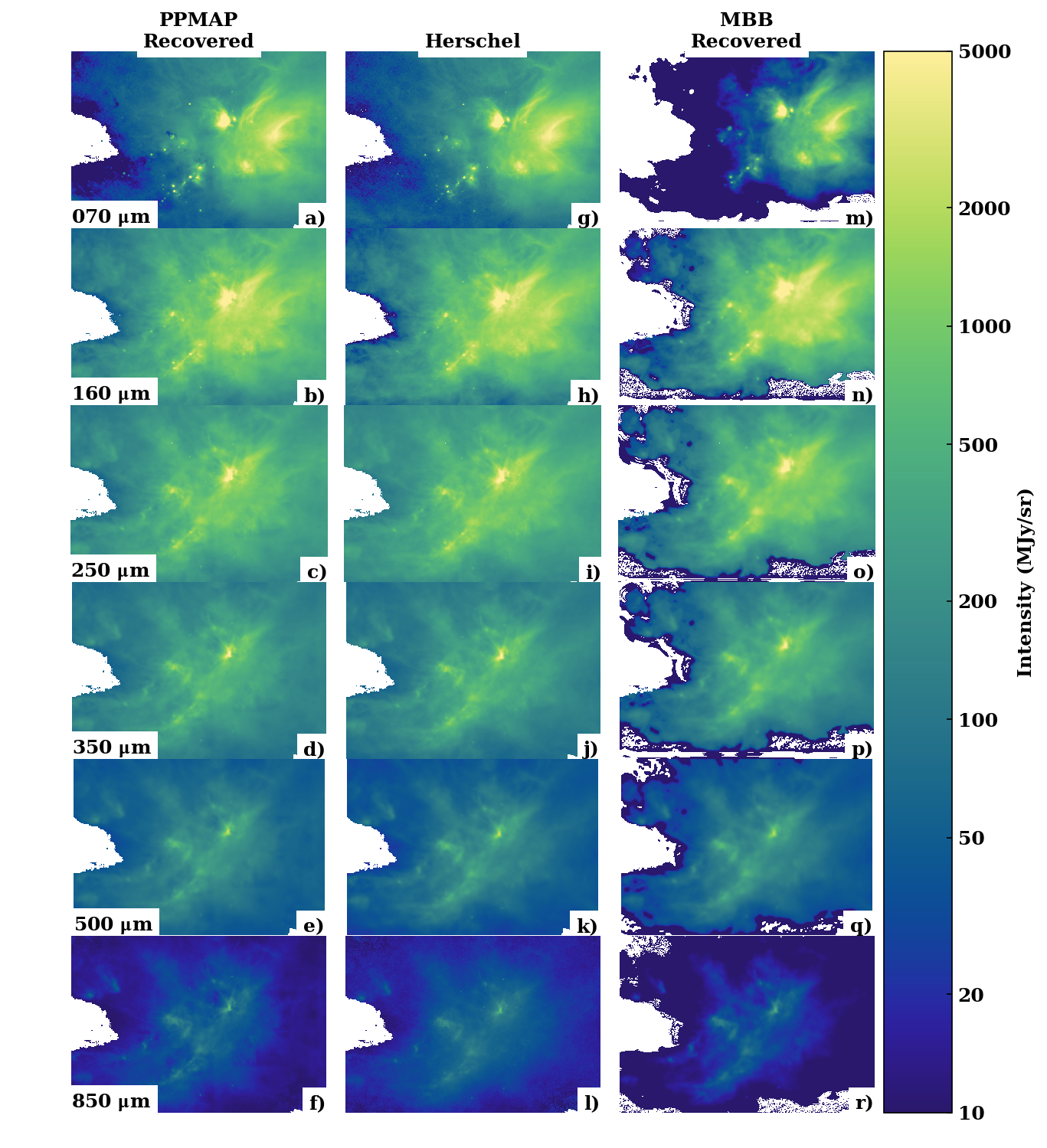}
    \caption{A montage of images of L1688. The central column shows the original images in the different {\it Herschel} and SCUBA2 wavebands. The lefthand column shows synthetic images derived from the PPMAP estimates of $T-$ and $\beta-$differential optical depth (i.e. Eq. \ref{EQN:PPInt}). The righthand column shows synthetic images from the MBB estimates of ${\bar T}$ (i.e. Eq. \ref{EQN:MBBInt}).}
    \label{FIG:L1688synthetic}
\end{figure*}

\section{PPMAP Data Products}\label{sec:ppmap-prod}

\subsection{Basic data products}

The four dimensional data-hypercubes produced by the PPMAP algorithm are hard to visualise. However, given the expectation values, $\Delta^2\tau_{{0:k\ell}}$, we can marginalise out one of the line of sight dimensions, $\beta_k$ or $T_\ell$, to produce a three dimensional data-cube. These data-cubes are analogous to position-position-velocity data-cubes derived from spectral line analysis, but with the velocity dimension replaced with either opacity index $\beta$ or dust temperature $T$, and with the integrated intensity replaced with optical depth. The results can then be displayed as a sequence of 2D images. 

By marginalising out the $\beta$ dimension, we obtain
\begin{equation}
    \label{EQN:3DTemp}
    \Delta\tau_{{0:\ell}} = \sum\limits_{k=1}^{k=3}\left\{\Delta^{\!2}\tau_{{0:k\ell}}\right\},
\end{equation}
where $\Delta\tau_{{0:\ell}}$ is the contribution to the total line of sight opacity from dust at temperature $T_\ell$. 

Similarly, by marginalising out the $T$ dimension, we obtain
\begin{equation}
    \label{EQN:3DBeta}
    \Delta\tau_{{0:k}} = \sum\limits_{\ell=1}^{\ell=12}\left\{\Delta^{\!2}\tau_{{0:k\ell}}\right\},
\end{equation}
where $\Delta\tau_{{0:k}}$ is the contribution to the total line of sight opacity from dust with opacity index $\beta_k$.

A two dimensional map of the expectation value for the total optical depth is obtained by marginalising out both $\beta$ and $T$,
\begin{equation}
    \label{EQN:cdens}
    \tau_{0} = \sum\limits_{\ell=1}^{\ell=12} \sum\limits_{k=1}^{k=3} \left\{\Delta^{\!2}\tau_{{0:k\ell}}\right\}.
\end{equation}

The corresponding 2D map of the uncertainty on the expectation value is obtained by summing the individual contributions from $\beta$ and $T$ in quadrature
\begin{equation}
    \label{EQN:uncert}
    \sigma_{0} = \left( \sum\limits_{\ell=1}^{\ell=12} \sum\limits_{k=1}^{k=3} \left\{\left(\Delta^{\!2}\sigma_{{0:k\ell}}\right)^2\right\} \vphantom{\sum\limits_{\ell=1}^{\ell=12}} \right)^{1/2}. 
\end{equation}

By combining Eqs. (\ref{EQN:cdens}) and (\ref{EQN:uncert}), we produce a map of the Point Process Statistical Degeneracy,
\begin{equation}
    \label{EQN:PPSD}
    \mathrm{PPSD} = \tau_0/\sigma_0\,.
\end{equation}
PPSD is a measure of how statistically significant the PPMAP optical depth estimates are, on a pixel-by-pixel basis, and is analogous to a signal-to-noise ratio. 

Additionally, we can obtain maps of the mean dust temperature, 
\begin{equation}
    \label{EQN:T_bar}
    \bar{T} = \frac{1}{\tau_{0}} \sum\limits_{\ell=1}^{\ell=12} \left\{ T_\ell \;\Delta\tau_{{0:\ell}} \vphantom{X^Y_Z} \right\},
\end{equation}
and the mean dust opacity index, 
\begin{equation}
    \label{EQN:beta_bar}
    \bar{\beta} = \frac{1}{\tau_{0}} \sum\limits_{k=1}^{k=3} \left\{ \beta_k\; \Delta\tau_{{0:k}} \vphantom{X^Y_Z} \right\},
\end{equation}
along each line of sight.

As discussed in \S \ref{SEC:ConvFact}, while PPMAP estimates dust opacity, we present our results in terms of the surface density of gas and dust, $\Sigma$, using Eq. (\ref{EQN:tau2Sigma}). We note, however, that the observations, and thus the PPMAP models, trace dust rather than directly tracing gas.

Note that on Figs. \ref{fig:obsChart} through \ref{fig:L1689ContourCore}, the scales of the colour bars for intensity, surface-density and dust temperature are logarithmic. This means that, in order to draw out the important features, the range of values represented must have a finite minimum and a finite maximum, and cannot extend down to zero. On the plots, but not in the analysis, positive values that fall below this minimum value are increased to the minimum. In pixels where there is no value (or in the case of fluxes, especially those from near the edges of the SCUBA2 pointings, negative values) these values are set to zero and the pixels are white.

\subsection{L1688}\label{sec:ppmap-prod:L1688}

Fig. \ref{fig:L1688TempGrid} shows the temperature slices for the L1688 sub-region of Ophiuchus at nine contiguous temperatures ($T_3 = \SI{10.0}{\kelvin}$, $T_4 = \SI{12.0}{\kelvin}$, $T_5 = \SI{14.3}{\kelvin}$, $T_6 = \SI{17.1}{\kelvin}$, $T_7 = \SI{20.5}{\kelvin}$, $T_8 = \SI{24.5}{\kelvin}$, $T_9 = \SI{29.2}{\kelvin}$, $T_{10} = \SI{35.0}{\kelvin}$, $T_{11} = \SI{41.8}{\kelvin}$). On large scales, there is a temperature gradient, from colder material ($\la 14\,{\rm K}$) towards the east of the sub-region, to warmer material ($\ga 17\,{\rm K}$) through the centre and towards the western edge. Several filamentary structures can be seen throughout the sub-region. While these structures subscribe to the large scale temperature gradient, it is also generally true that any single structure which dominates over its local background in lower-temperatures slices appears to fade relative to that background as we progress to higher temperatures, until only the local background is present. This indicates that the dense, filamentary structures are typically somewhat colder than their surroundings. Two regions of diffuse, hot ($\ga 29\,{\rm K}$) dust surround the pre-main sequence objects S1 and HD147889 (red star markers), suggesting significant local heating by these young B stars.

Fig. \ref{fig:L1688BetaGrid} shows the opacity-index slices for the L1688 sub-region for all three discrete values of $\beta$. The $\beta_1\!=\!1.0$ slice is dominated by the structures associated with material in the coldest (\SI{10.0}{\kelvin}) slice of Fig. \ref{fig:L1688TempGrid}. Warmer structures, and the diffuse background environment, are better traced by material with higher opacity index $\beta\ga 1.5$. This suggests that there is a change in the optical properties of dust in colder environments, similar to the trend found by \citet{Howard2019} in the Taurus molecular cloud.

Fig. \ref{fig:L1688Cdens} shows the estimated total surface density, $\Sigma$,  derived from Eqs. (\ref{EQN:tau2Sigma}) and (\ref{EQN:cdens}). The filamentary structures observed in Fig. \ref{fig:L1688TempGrid} are clearly visible in this map of total column density, and can also be cross referenced with the dense clumps (Rho Oph A through Rho Oph F) identified by \citet{Loren1990} from observations of DCO$^+$. The approximate locations of these clumps are shown by the letter identifiers on Fig. \ref{fig:L1688Cdens}. The low density regions surrounding S1 and HD147889 are also clearly visible. The map of the PPSD for L1688 is given in Appendix \ref{apx:PPSD}.

Figs. \ref{fig:L1688TB}a and b show the mean line-of-sight dust temperature (Eq. \ref{EQN:T_bar}) and the mean line-of-sight dust opacity index (Eq. \ref{EQN:beta_bar}) for the L1688 sub-region. Fig. \ref{fig:L1688TB}a clearly shows the temperature gradient across the region, as well as the additional heating from S1 and HD147889. The mean temperatures of the dust in the filamentary structures seen in Fig. \ref{fig:L1688Cdens} do not generally differ greatly from those in their local surroundings, which makes them difficult to identify clearly on Fig. \ref{fig:L1688TB}a. The exceptions are the structures associated with Rho Oph B, and, to a lesser extent, Rho Oph D, which are noticeably cooler.

The map of mean dust opacity index (Fig. \ref{fig:L1688TB}b) does not trace the fine structure of the region as clearly as the maps of surface density (Fig. \ref{fig:L1688Cdens}) or mean dust temperature (Fig. \ref{fig:L1688TB}a). However, the dust in regions with higher surface density on Fig. \ref{fig:L1688Cdens} tends to have $\bar{\beta} \la 1.3$, while the dust in regions with lower surface density tends to have $\bar{\beta} \ga 1.7$. A similar trend of dense regions harbouring dust with a lower opacity index than the more diffuse background material was observed in Taurus by \citet{Howard2019}. Proximity to the B stars S1 and HD147889 does not appear to influence the opacity index greatly.

Fig. \ref{FIG:L1688synthetic} shows a montage of images in all the wavebands used (i.e. the five {\it Herschel} bands at \SI{70}{\micro \meter}, \SI{160}{\micro \meter}, \SI{250}{\micro \meter}, \SI{350}{\micro \meter}, \SI{500}{\micro \meter}, and the SCUBA2 band at \SI{850}{\micro \meter}). The central column shows the actual observed images. The lefthand column shows synthetic images generated from the PPMAP estimates of the $T$- and $\beta$-differential optical depth contributions, $\Delta^2\tau_{0:k\ell}$, on the line of sight through each pixel (i.e Eq. \ref{EQN:PPInt}). The righthand column shows synthetic images generated from the MBB estimates of the average optical depth, ${\bar\tau}_0$, and temperature, ${\bar T}$, on the line of sight through each pixel (i.e. Eq. \ref{EQN:MBBInt}). In all cases the images are at the native resolution of the corresponding {\it Herschel} image. Where the resolution of the {\sc ppmap} or MBB results is coarser than the corresponding {\it Herschel} image, values at the native resolution have been generated by interpolation.

To evaluate the fidelity of the results, we compute a goodness-of-fit metric, 
\begin{equation}\label{EQN:GoF}
{\cal G}_{_{\rm PROC:WB}}\!\!=\!\!\left\{\frac{\sum\limits_{_{\rm PIXELS}}\left\{I_{_{\rm TRUE:WB}}^{-1}\left[I_{_{\rm PROC:WB}}-I_{_{\rm TRUE:WB}}\right]^2\right\}}{\sum\limits_{_{\rm PIXELS}}\left\{I_{_{\rm TRUE:WB}}\right\}}\right\}^{1/2}\!,\hspace{0.8cm}
\end{equation}
for each synthetic image \citep[cf.][]{Howard2019}. In Eq. \ref{EQN:GoF}, the `$I$'s are the intensities in individual pixels, the summations are over all pixels, the subscript `{\sc proc}' stands for the procedure used to generate the synthetic intensities (i.e. PPMAP or MBB), the subscript `{\sc true}' stands for the actual observations, and the subscript `{\sc wb}' stands for the waveband under consideration. Values of ${\cal G}$ are summarised in Table \ref{TAB:GoF}. We see that in all but one of the wavebands the PPMAP procedure produces a significantly better fit to the true observations than the MBB procedure. Moreover, in the one waveband where MBB produces the better fit (\SI{500}{\micro \meter}), the difference in ${\cal G}$ values is small. The goodness-of-fit values, ${\cal G}$, obtained by {\sc ppmap} for Ophiuchus are significantly worse (higher) than those reported for Taurus in \citet{Howard2019}, because the Ophiuchus analysis is compromised by the very noisy edges of the mapped regions.

\begin{table}
\centering
\caption{Values of the goodness-of-fit metric, ${\cal G}$ (Eq. \ref{EQN:GoF}), for synthetic maps of the two subregions (L1688 and L1689), based on the PPMAP procedure and the MBB procedure.}
\begin{tabular}{lcccccc}\hline
{\sc Subregion} & & \multicolumn{2}{c}{L1688} & & \multicolumn{2}{c}{L1689} \\
{\sc Procdedure} &\hspace{0.3cm} & PPMAP & MBB & \hspace{0.3cm} & PPMAP & MBB \\\hline
{\sc Waveband} & & & & & & \\
\SI{70}{\micro \meter} & & 0.21 & 1.21 & & 0.57 & 0.97 \\
\SI{160}{\micro \meter} & & 0.29 & 0.66 & & 0.50 & 0.48 \\
\SI{250}{\micro \meter} & & 0.19 & 0.57 & & 0.18 & 0.57 \\
\SI{350}{\micro \meter} & & 0.20 & 0.55 & & 0.17 & 0.43 \\
\SI{500}{\micro \meter} & & 0.28 & 0.21 & & 0.28 & 0.30 \\
\SI{850}{\micro \meter} & & 0.32 & 0.54 & & 0.32 & 0.56 \\\hline
\end{tabular}
\label{TAB:GoF}
\end{table}

\subsection{L1689}\label{sec:ppmap-prod:L1689}

\begin{figure*}
    \centering
    \includegraphics[width = 1.0\textwidth]{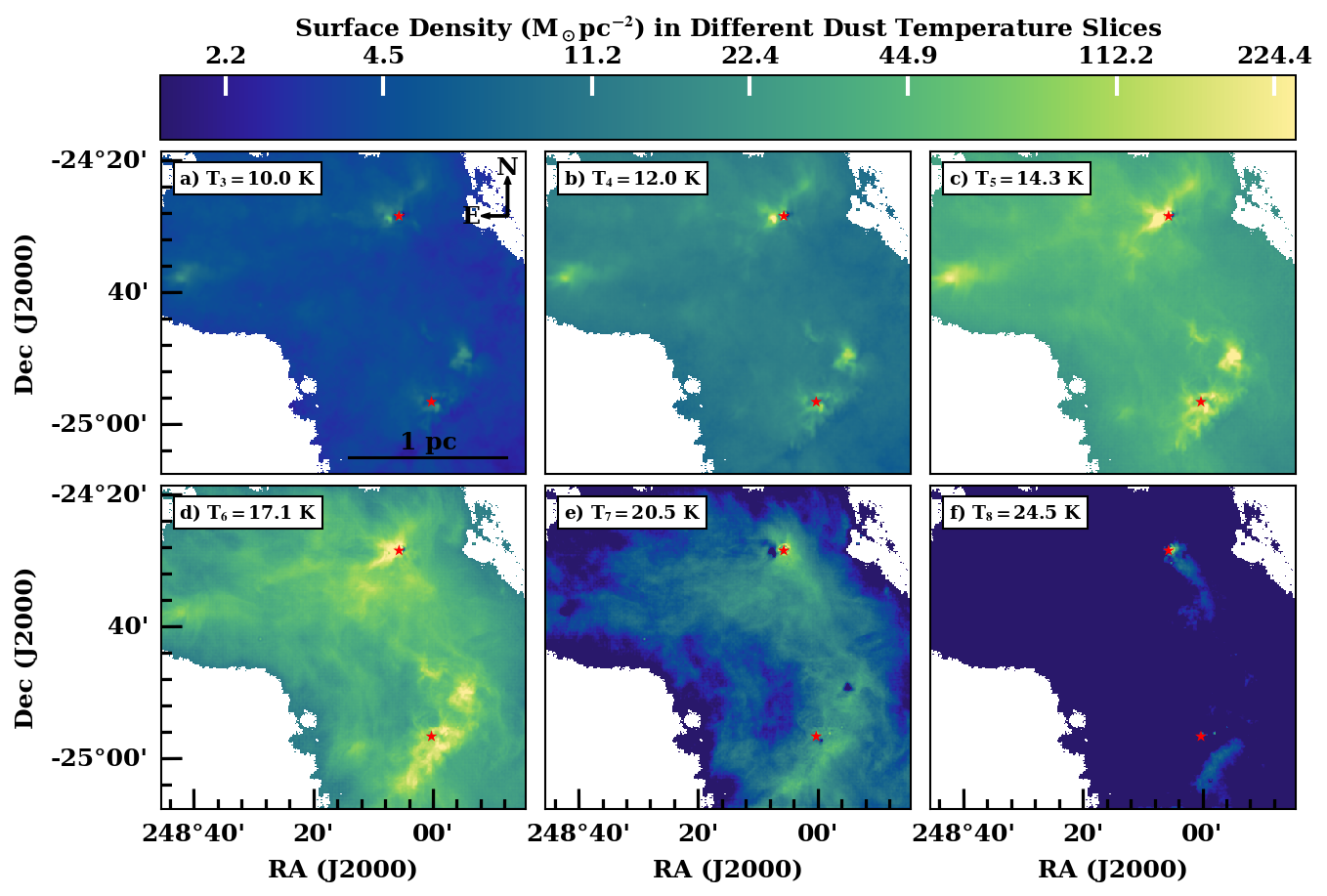}
    \caption{Six contiguous PPMAP temperature slices for the L1689 sub-region, showing the surface density, $\Sigma$, as traced by dust having temperature represented by the value marked in the top left. The red stars indicate the positions of L1689-IRS6 and 16293-2422.}
    \label{fig:L1689TempGrid}
\end{figure*}

\begin{figure*}
    \centering
    \includegraphics[width = 1.0\textwidth]{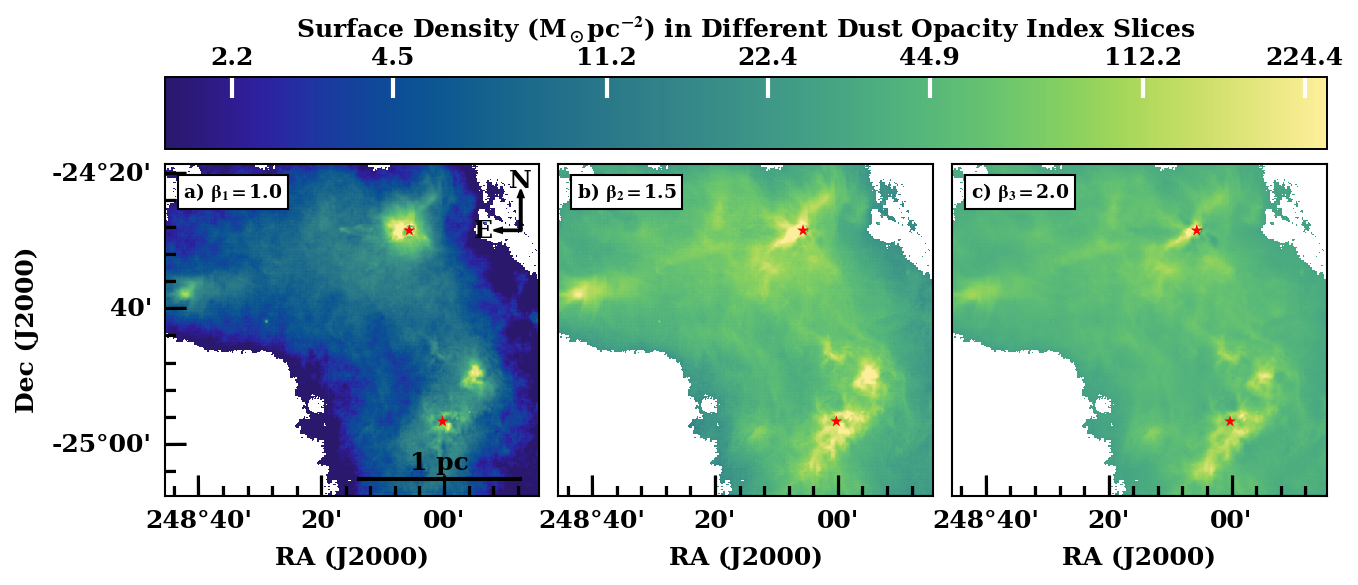}
    \caption{Three contiguous PPMAP opacity-index slices for the L1689 sub-region, showing the surface density, $\Sigma$, as traced by dust having opacity index represented by the value marked in the top left. The red stars indicate the positions of L1689-IRS6 and 16293-2422.}
    \label{fig:L1689BetaGrid}
\end{figure*}

\begin{figure*}
    \centering
    \includegraphics[width = 1.0\textwidth]{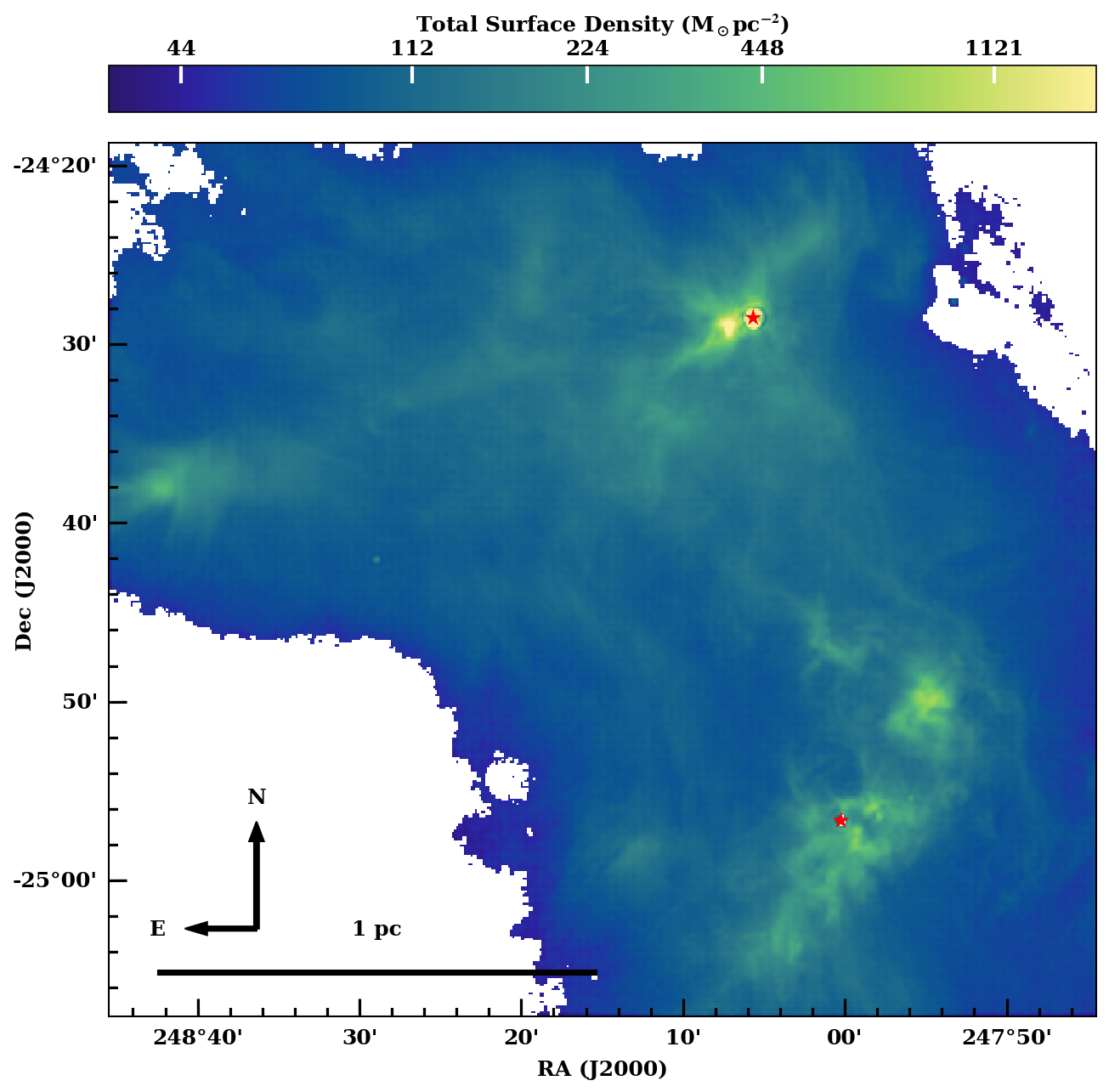}
    \caption{The estimated total surface density, $\Sigma$, for the L1689 sub-region. The red stars indicate the positions of the L1689-IRS6 and 16293-2422 protostellar systems.}
    \label{fig:L1689Cdens}
\end{figure*}

\begin{figure*}
    \centering
    \includegraphics[height = 0.95\textheight]{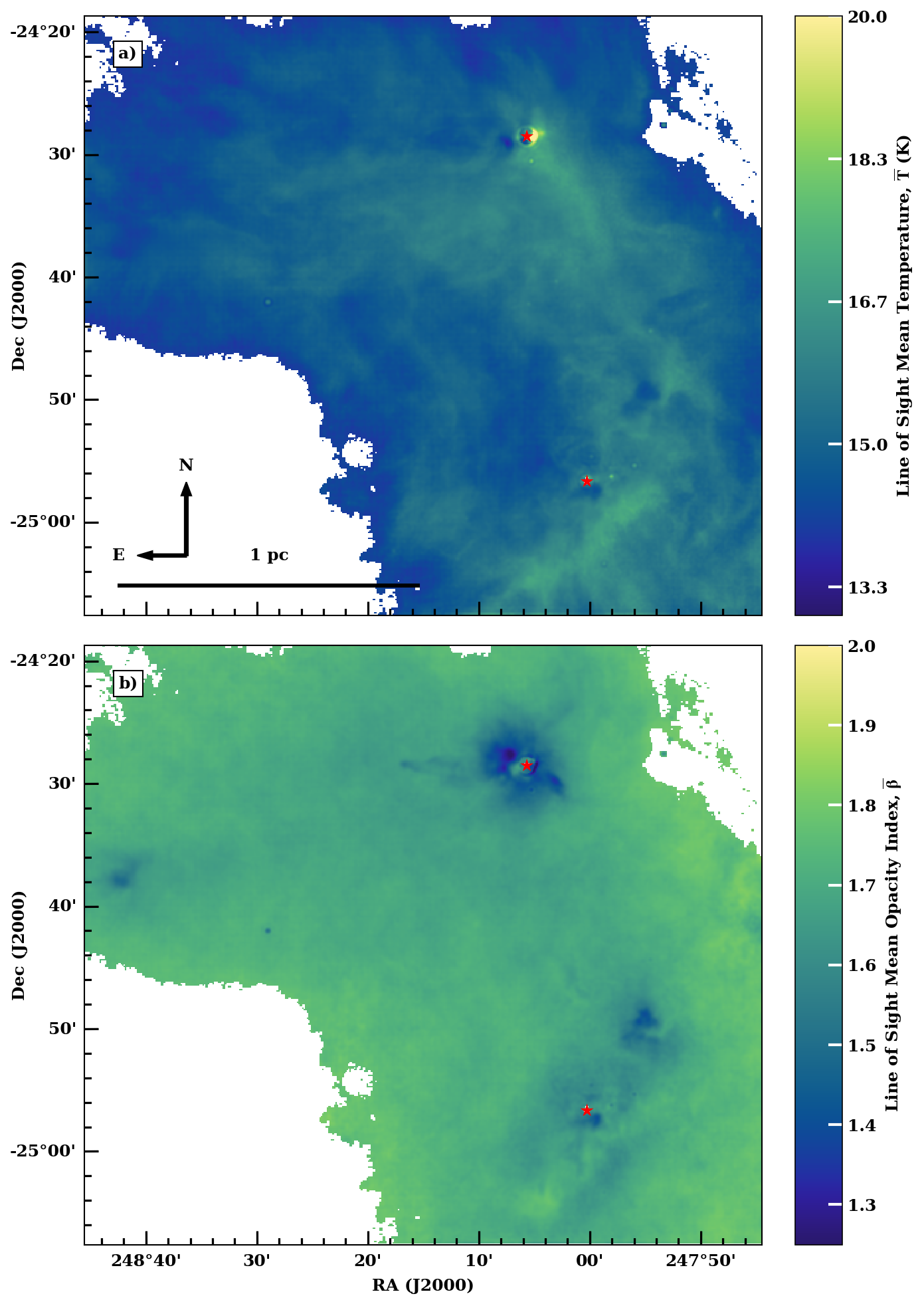}
    \caption{Maps of (a) the mean line-of-sight dust temperature (Eq. \ref{EQN:T_bar}), and (b) the mean line-of-sight dust opacity index (Eq. \ref{EQN:beta_bar}), for the L1689 sub-region. The red stars indicate the positions of the L1689-IRS6 and 16293-2422 pre-stellar systems.}
    \label{fig:L1689TB}
\end{figure*}

\begin{figure*}
    \centering
    \includegraphics[height = 0.93\textheight]{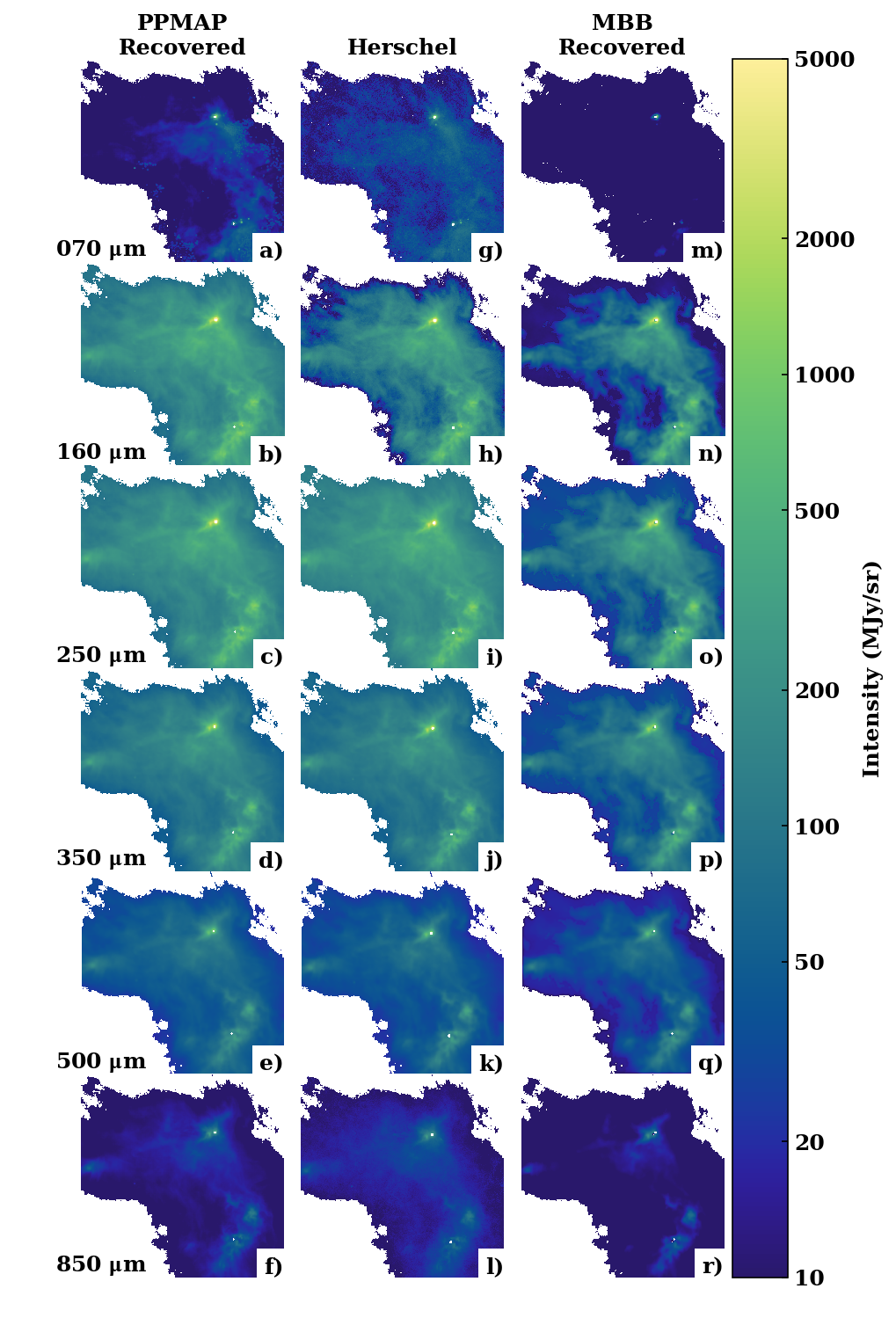}
    \caption{A montage of images of L1689. The central column shows the original images in the different {\it Herschel} and SCUBA2 wavebands. The lefthand column shows synthetic images derived from the PPMAP estimates of $T-$ and $\beta-$differential optical depth (i.e. Eq. \ref{EQN:PPInt}). The righthand column shows synthetic images from the MBB estimates of ${\bar T}$ (i.e. Eq. \ref{EQN:MBBInt}).}
    \label{FIG:L1689synthetic}
\end{figure*}

Fig. \ref{fig:L1689TempGrid} shows temperature slices for the L1689 sub-region of Ophiuchus at six contiguous temperatures ($T_3 = \SI{10.0}{\kelvin}$, $T_4 = \SI{12.0}{\kelvin}$, $T_5 = \SI{14.3}{\kelvin}$, $T_6 = \SI{17.1}{\kelvin}$, $T_7 = \SI{20.5}{\kelvin}$ and $T_8 = \SI{24.5}{\kelvin}$). There are three dense, compact, and seemingly isolated structures in the temperature slices with $T \leq \SI{12.0}{\kelvin}$, to the south, north and north east of the sub-region; these structures correspond to the dense clumps identified by \citet{Loren1990}, respectively L1689-South, L1689-North and L1689-East. The slices at temperatures \SI{\geq 14.3}{\kelvin} reveal a network of interconnecting filamentary structures between these clumps, which are not prominent at lower temperatures. In the \SI{\geq 17.1}{\kelvin} slices, L1689-East is less visible than the other two clumps, indicating that it is not as warm. Overall the dust in L1689 is colder than that in L1688, with only two small ridges of material at \SI{24.5}{\kelvin}; this material is located near the masked protostellar systems, L1689-IRS6 and 16293-2422, and probably experiences additional local heating from these sources. The overall temperature difference between L1689 and L1688 is probably attributable to the earlier evolutionary state of L1689-IRS6 and 16293-2422 (as compared with S1 and HD147889), and/or to the greater distance of L1689 from the Upper Sco OB association.

Fig. \ref{fig:L1689BetaGrid} shows the opacity-index slices for the L1689 sub-region for all three discrete values of $\beta$. As with L1688, a much larger proportion of the material in the $\beta = 1.0$ slice is found within the central regions of the dense, cold clumps than in the surrounding medium. This diffuse medium is better traced by material with a higher opacity index.

Fig. \ref{fig:L1689Cdens} shows the estimated total surface density, $\Sigma$, derived from Eqns.  (\ref{EQN:tau2Sigma}) and (\ref{EQN:cdens}). The L1689-South, L1689-North and L1689-East clumps are still clearly visible, although the network of diffuse interconnecting filamentary structures is less apparent than in Fig. \ref{fig:L1689TempGrid}. The map of the PPSD for L1689 is given in Appendix \ref{apx:PPSD}.

Figs. \ref{fig:L1689TB}a and b show the mean line-of-sight dust temperature (Eq. \ref{EQN:T_bar}) and the mean line-of-sight dust opacity index (Eq. \ref{EQN:beta_bar}), for the L1689 sub-region. On Fig. \ref{fig:L1689TB}a, the dense clumps (L1689-South, L1689-North and L1689-East) and the interconnecting filamentary structures appear to be somewhat warmer ($\sim 17\,{\rm K}$) than the more diffuse surroundings ($\sim 14\,{\rm K}$). There are several possible explanations for this seemingly paradoxical result, all of which relate to the fact that the line-of-sight averaging combines contributions from different regions along the line of sight with different radiation fields, and also contributions from different types of dust which may co-exist in the same volume. One explanation is that the lines of sight away from the dense clumps have significant contributions from regions that are not close to Ophiuchus, and as a consequence have a weaker background radiation field and cooler dust; a threefold reduction in the ambient radiation field would reduce the equilibrium temperature of a typical dust grain (of the same type) from $\sim\!17\,{\rm K}$ to $\sim\!14\,{\rm K}$. A second explanation is that there is significant extra dust heating within the clumps, due to embedded or nearby sources (including L1689-IRS6 and 16293-2422); on Fig. \ref{fig:L1689TempGrid}f (the $T_8\!=\!24.5\,{\rm K}$ panel) there is evidence for local heating fronts to the south-west of L1689-North and to the south-west of L1689-South.

Fig. \ref{fig:L1689TB}b shows the mean line-of-sight opacity index for the L1689 sub-region. As in L1688, the diffuse dust has a relatively high mean opacity index, $\bar{\beta} \ga 1.7$, while the dust in the denser regions has a lower value, $\bar{\beta} \la 1.4$.

Fig. \ref{FIG:L1689synthetic} shows a montage of images of L1689, equivalent to those in Fig. \ref{FIG:L1688synthetic}. The goodness-of-fit metric, ${\cal G}$ (see Eq. \ref{EQN:GoF}) for each synthetic image is given in Table \ref{TAB:GoF}. Again, in all but one waveband the PPMAP procedure produces a significantly better fit than the MBB procedure. In the one waveband where this is not the case (\SI{160}{\micro \meter}), the PACS images are very noisy in the low-intensity regions and the ${\cal G}$ values are very similar.

\section{The mass distribution in L1688 and L1689}\label{sec:mass}

The {\sc ppmap}  pixels have angular size $\Delta\Omega_{\rm pixel}\!=\!49\,{\rm arcsec^2}$, and Ophiuchus is assumed to be at $D\!=\!140\,{\rm pc}$. Hence, from Eq. (\ref{EQN:massFromCdens}), the mass in a single pixel is
\begin{equation}
\Delta M_{\rm pixel}\sim 1.1\,{\rm M}_\odot\;\tau_0\,.
\end{equation}
Since the uncertainties in this conversion are dominated by the mass opacity ($\propto\kappa_0^{-1}$) and the distance ($\propto D^2$), we do not quote the {\sc ppmap} uncertainties (which are much smaller, see Appendix \ref{apx:PPSD}) and we give all masses and surface densities to two significant figures. Even this is unduly optimistic for absolute values, but relative values should be more reliable. The results are summarised in Table \ref{tab:mass}.

Summing the pixels on Fig. \ref{fig:L1688Cdens}, the estimated mass and mean surface density of the L1688 sub-region (the region outlined in black on Fig. \ref{fig:obsChart}) are $650\,{\rm M}_\odot$ and $110\,{\rm M_\odot\,pc^{-2}}$. The corresponding quantities for L1689 (Fig. \ref{fig:L1689Cdens}) are $400\,{\rm M}_\odot$ and $66\,{\rm M_\odot\,pc^{-2}}$.

\begin{figure*}
    \centering
    \includegraphics[width = 1.0\textwidth]{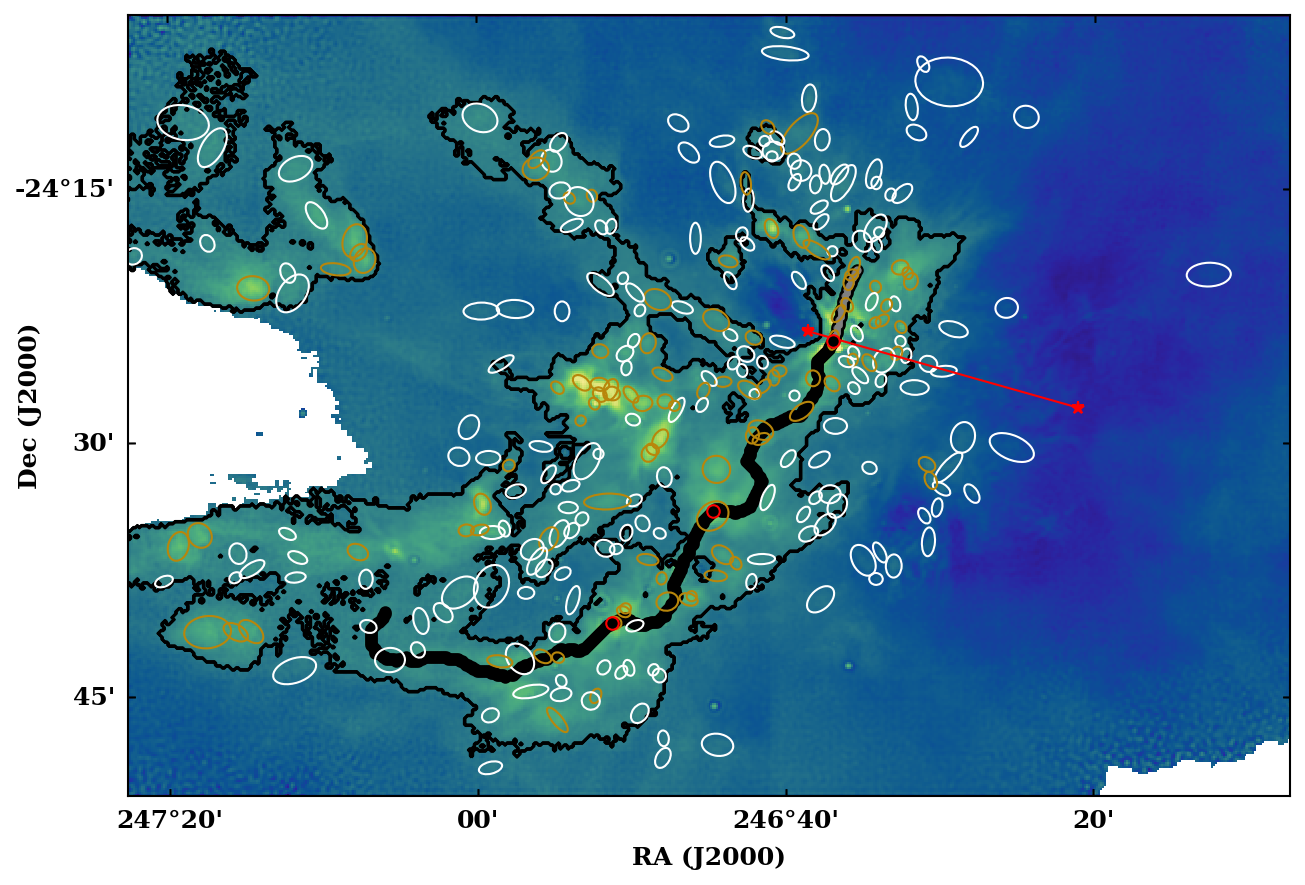}
\caption{The total surface density, $\Sigma$, for L1688 (as Fig. \ref{fig:L1688Cdens}) with the High Efficiency Regions (HERs) outlined in {\bf black}, and the starless and pre-stellar cores identified by \citet{Ladjelate2020} shown as, respectively, white and gold ellipses. The red stars mark the locations of the B stars S1 and HD147889, and the red line connecting them (the \AoI) is discussed further in Section \ref{sec:f1feedback}. Filament \texttt{f1} is traced with black filled circles. The red outlined circles represent local column density maxima along the filament spine.}
    \label{fig:L1688ContourCore}
\end{figure*}

\begin{figure*}
    \centering
    \includegraphics[width = 1.0\textwidth]{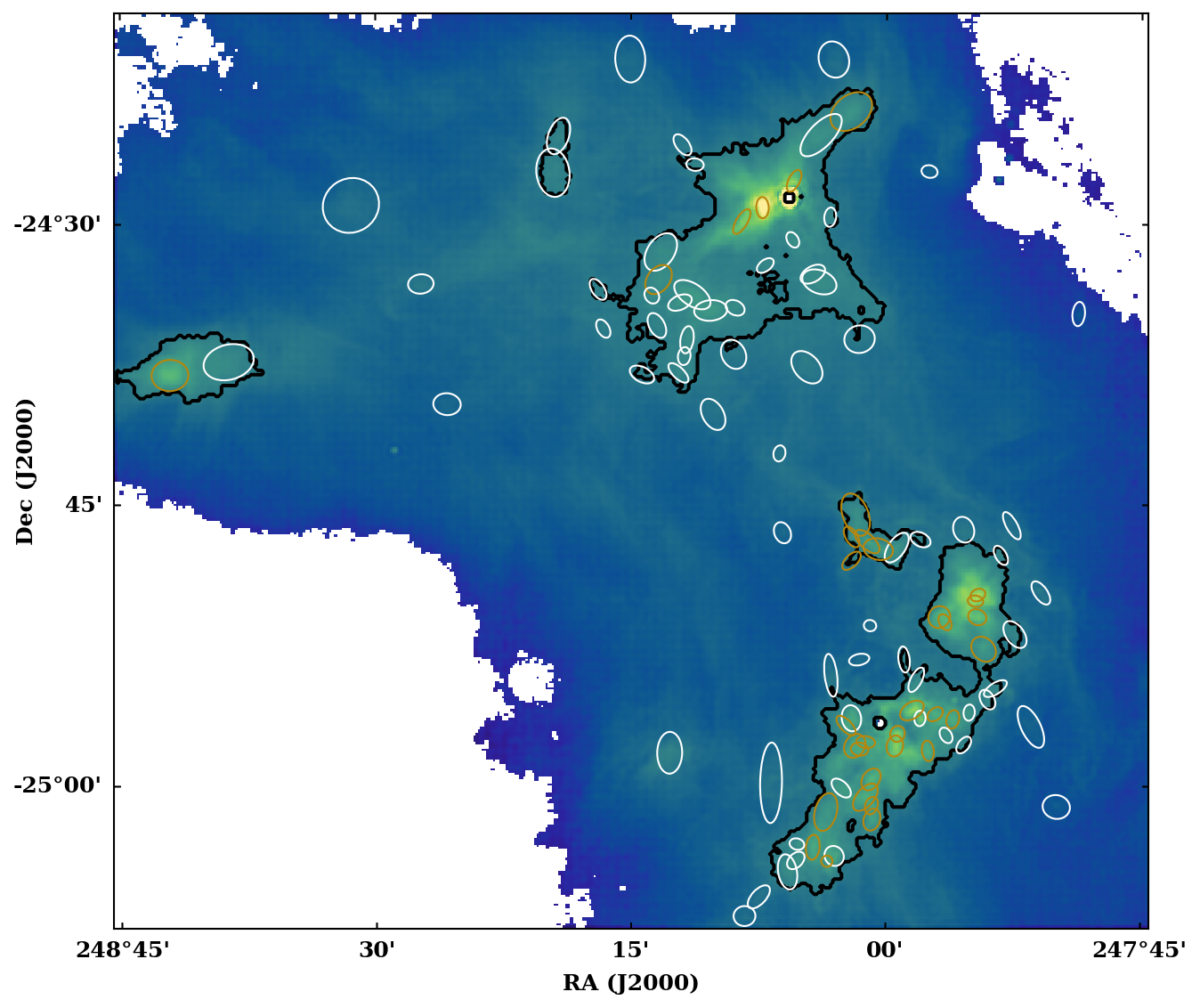}
    \caption{The total surface density, $\Sigma$, for L1689 (as Fig. \ref{fig:L1689Cdens}) with the High Efficiency Regions (HERs) outlined in {\bf black}, and the starless and pre-stellar cores identified by \citet{Ladjelate2020} shown as, respectively, white and gold ellipses.}
    \label{fig:L1689ContourCore}
\end{figure*}

However, the rectangular boundaries defining the sub-regions are somewhat arbitrary. In order to make our analysis more objective, we focus on the {\it High Efficiency Regions} (HERs) of L1688 and L1689. The HERs are the regions with surface density $\Sigma>\Sigma_{_{\rm T}}\simeq 160\,{\rm M_\odot\,pc^{-2}}$ (and hence $A_{_{\rm V}}>A_{_{\rm T}}\simeq 7\,{\rm mag}$). Several studies \citep[e.g.][]{Andre2010, Lada2010, Konyves2015} have found that this threshold column density marks an abrupt transition to high star formation efficency. Figs. \ref{fig:L1688ContourCore} and \ref{fig:L1689ContourCore} show maps of $\Sigma$ (the same as Figs. \ref{fig:L1688Cdens} and \ref{fig:L1689Cdens}), with the boundaries of the HERs marked in black. The mass, area and mean surface density of the L1688 HER are $290\,{\rm M_\odot}$, $1.2\,{\rm pc^2}$ and $240\,{\rm M_\odot\,pc^{-2}}$. For the L1689 HER, these quantities are $89\,{\rm M_\odot}$, $0.37\,{\rm pc^2}$ and $240\,{\rm M_\odot\,pc^{-2}}$. The masses derived here are  lower than those obtained by \citet{Ladjelate2020} because their MBB fit returns a single, flux-averaged mean temperature along each line of sight. Consequently it overestimates the contribution from hotter than average dust, and underestimates the contribution from colder than average dust. \citet{Marsh2015} have shown that, when using a flux-averaged mean, the overestimate usually dominates, and therefore total masses are often overestimated -- sometimes by as much as $100\%$, but here by $\sim 50\%$.

To analyse the population of starless and pre-stellar cores in L1688 and L1689, we use the catalogue of \citet{Ladjelate2020}, which identifies 513 cores in Ophiuchus, and fits them with ellipses.\footnote{The convention we have adopted is that `starless cores' are cores that do not contain stars and appear not to be gravitationally bound; they are therefore likely to disperse unless their circumstances change, for example some other force acts to contain them or they increase their mass by accretion or merger. Conversely, `pre-stellar cores' are cores that do not contain stars but appear to be gravitational bound; they are therefore expected to collapse and form a single star, or a multiple system, or a small sub-cluster of stars.} 49 of these cores are identified as protostellar. Consequently they are optically thick, and {\sc ppmap} cannot fit them. Of the remaining 464 starless and pre-stellar cores, 288 lie in the the L1688 sub-region, and 101 in the L1689 sub-region. The elliptical fits for these cores are shown in white on Figs. \ref{fig:L1688ContourCore} and \ref{fig:L1689ContourCore}. 

We estimate the masses of the starless and pre-stellar cores by applying the \texttt{photutils} Python package \citep{Bradley2016} to the maps of total \si{\htwo} column density, using elliptical apertures corresponding to the fits given by \citet{Ladjelate2020}. The 288 cores in the L1688 sub-region have a combined mass of \SI[separate-uncertainty=true]{100}{\msol}, and thus account for $16\%$ of the total mass of the sub-region. The 101 cores in the L1689 sub-region have a combined mass of \SI[separate-uncertainty=true]{41 \pm 0.1}{\msol}, making up $10\%$ of the total mass of the sub-region. 

Figs. \ref{fig:L1688ContourCore} and \ref{fig:L1689ContourCore} show that the majority of the cores lie within the HERs. By excluding the cores that lie outside of the HERs in L1688, we find the remaining 219 cores have a total mass of \SI[separate-uncertainty=true]{84}{\msol}, which accounts for $29\%$ of the HER mass. In L1689, the 81 cores found within the HER contain \SI[separate-uncertainty=true]{32}{\msol}, or $36\%$ of the HER mass. These percentages are higher than those reported by \citet{Ladjelate2020}, firstly because the masses we derive for the HERs are lower, as explained above, and secondly because we have summed the masses of all cores, starless and prestellar.

Thus, the fraction of the total mass in the L1688 sub-region that is in cores is $50\%$ times higher than in the L1689 sub-region. In contrast, the fraction of the total mass in the L1688 HER that is in cores is much closer to -- and actually $20\%$ {\it lower than} --  the fraction in the L1689 HER. This suggests that the core formation efficiency is more strongly dependent on local surface density than total cloud mass. 

Essentially the same trends are found if we use the core masses obtained by \citet{Ladjelate2020} using modified black body fits to the \textit{Herschel} observations, or the core masses obtained by  \citet{Nutter2006} using both a different method of mass estimation and a different core catalogue.

\begin{figure}
    \centering
    \includegraphics[width = \linewidth]{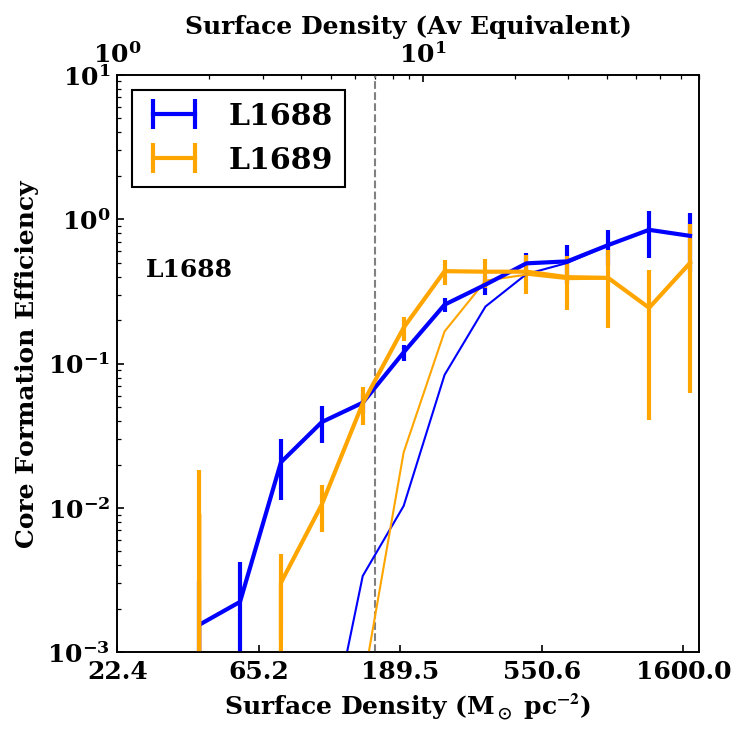}
    \caption{Log-log plot of the core formation efficiency, $\eta$, as a function of the threshold surface density, $\Sigma$ (Eq. \ref{EQN:CFE}). {\it Blue curves:} L1688. {\it Orange curves:} L1689. {\it Thick curves:} starless and prestellar cores together. {\it Thin curves:} prestellar cores alone. The vertical dashed line marks the threshold, $\Sigma_{_{\rm T}}\sim 160\,\rm{M_\odot\,pc^{-2}}$, estimated by \citet{Konyves2015}; these results suggest that in Ophiuchus the threshold may be a little higher, $\Sigma_{_{\rm T}}\sim 270\,\rm{M_\odot\,pc^{-2}}$.}
    \label{fig:CFE}
\end{figure}

We confirm the existence of a threshold surface density, $\Sigma_{_{\rm T}}$, for efficient star formation by repeating the above analysis with different values for $\Sigma$. Following the procedure outlined in \citet{Konyves2015}, we vary $\Sigma$ and compute the efficiency,
\begin{equation}
\label{EQN:CFE}
\eta(\Sigma) = \frac{M_{_{\rm CORES}}(\Sigma)}{M_{_{\rm TOTAL}}(\Sigma)}\,,
\end{equation}
where $M_{_{\rm CORES}}(\Sigma)$ is the net mass of cores inside the iso-contour at $\Sigma$ and $M_{_{\rm TOTAL}}(\Sigma)$ is the total mass of everything (cores and intercore material) inside the iso-contour at $\Sigma$. Fig. \ref{fig:CFE} shows $\eta(\Sigma)$, on a log-log scale, for the L1688 sub-region (blue) and the L1689 sub-region (orange). The thin curves are for prestellar cores alone, and the thick curves are for starless and prestellar cores combined. These plots support the suggestion that there is a threshold surface-density above which the conversion of interstellar gas into prestellar cores is highly efficient, and below which it is very inefficient \citep{Konyves2015}. In Ophiuchus, the threshold appears to be at $\Sigma_{_{\rm T}}\sim 270(\pm 40)\,\rm{M_\odot\,pc^{-2}}$ (corresponding to $A_{_{\rm V}}\sim 12(\pm2$); this is the surface density above which prestellar cores contribute more mass than starless cores. It is somewhat higher than the $\Sigma_{_{\rm T}}\sim 160\,\rm{M_\odot\,pc^{-2}}$ (corresponding to $A_{_{\rm V}}\sim 7$) estimated by \citet{Konyves2015}. There may be local variations in the threshold, reflecting for example the intensity of the local radiation field. However, we should also keep in mind that our estimate is based on small-number statistics, and that there is uncertainty in how a prestellar core should be defined.

A related issue is what is likely to happen to the starless cores in the future. In the theory of turbulent fragmentation, it is inevitable that a significant fraction of starless cores are sterile, in the sense that they do not become prestellar, and therefore do not spawn stars \citep{Padoan2002}. These sterile cores either have too little mass to be strongly self-gravitating, or they are compressed too little (by the ram-pressure of the flows creating them and/or by he background gravitational field), and they disperse. As noted previously by \citet{Andre2010}, \citet{Andre2014} and \citet{Ladjelate2020}, in Ophiuchus -- and also in many other star formation regions -- the conditions for forming prestellar cores exist, almost exclusively, in filaments. Presumably the large-scale converging flows creating a filament deliver a reservoir of mass from which nascent cores can continue to accrete, a large ram-pressure to compress the gas, and a background gravitational field that will inhibit re-expansion. Conversely, cores that form in relative isolation, outside a filament, are much less likely to spawn stars; they have less ambient gas to accrete, lower external pressure, and no assistance from the collective gravitational field of a filament.

\begin{figure}
    \centering
    \includegraphics[width = \linewidth]{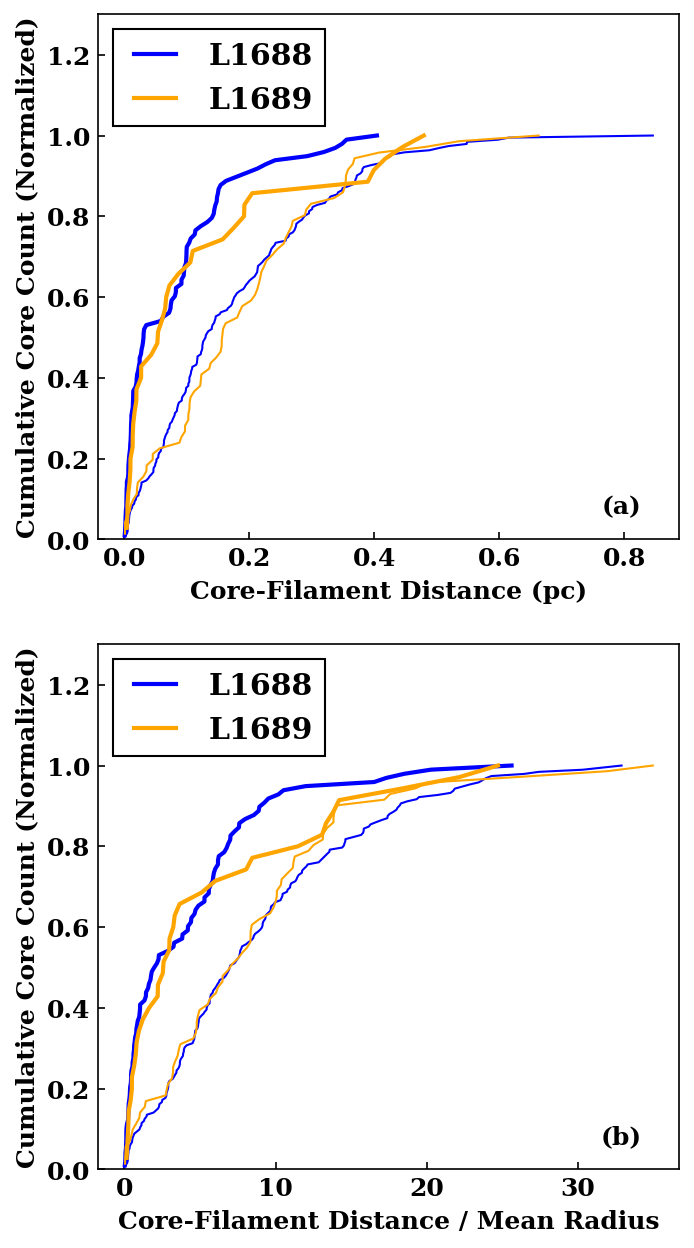}
    \caption{The cumulative distributions of (a) $D_{_{\rm C2F}}$ (the projected distance from a core to the nearest filament spine point) and (b) ${\cal R}=D_{_{\rm C2F}}/r_{_{\rm C}}$ (where $r_{_{\rm C}}$ is the mean radius of the core). {\it Blue curves:} L1688. {\it Orange curves:} L1689. {\it Thick curves:} prestellar cores. {\it Thin curves:} starless cores.}
    \label{FIG:DistOverRad}
\end{figure}

We suggest that many of the starless cores observed in Ophiuchus are indeed destined to disperse. On Fig. \ref{FIG:DistOverRad} we plot the cumulative distributions of $D_{_{\rm C2F}}$ (Fig. \ref{FIG:DistOverRad}a) and ${\cal R}=D_{_{\rm C2F}}/r_{_{\rm C}}$ (Fig. \ref{FIG:DistOverRad}b), where $D_{_{\rm C2F}}$ is the projected distance from a core to the nearest filament spine-point, $r_{_{\rm C}}=(ab)^{1/2}$ is the mean radius of the core, $a$ and $b$ are the semi-major and semi-minor axes of the fitted elliptical outline of the core from \citet{Ladjelate2020}. The thick curves represent the prestellar cores, and the thin curves represent the starless cores; blue curves represent L1688, and orange ones represent L1689. The timescale for a starless core to disperse is $t_{\rm disp}\la r_{_{\rm C}}/c_{_{\rm S}}$, where $c_{_{\rm S}}$ is the sound speed { (typically $0.25\pm 0.05\,\rm{km\,s^{-1}}$, corresponding to gas-kinetic temperatures between $\sim 11\,{\rm K}$ and $\sim 26\,{\rm K}$),} and `$\la$' is used because the core may already be dispersing. Conversely, the timescale for a starless core to shift to a filament is $t_{\rm shift}\ga D_{_{\rm C2F}}/{\cal M}c_{_{\rm S}}$, where ${\cal M}$ is the Mach Number of the core's bulk velocity and `$\ga$' is used because the bulk velocity must also be directed towards the nearest filament. A starless core is only likely to become prestellar if either a new filament forms around it, which seems unlikely, or it shifts to an existing filament on a timescale $t_{\rm shift}<t_{\rm disp}$, which requires ${\cal M}>{\cal R}$. The offset between the medians of the cumulative distributions on Fig. \ref{FIG:DistOverRad}b is ${\cal R}\sim 5$. This implies that very few of the starless cores are likely to reach one of the existing filaments before dispersing, unless they have bulk velocities with ${\cal M}>5$ ($v\!>\!1\,{\rm km\,s^{-1}}$), and this velocity is directed towards, or close to, the nearest part of that filament, which again seems unlikely.

To be more specific, we consider the starless cores on the $25^{\rm th}$ percentile of the distribution of $D_{_{\rm C2F}}$ values, which are at a projected distance $D_{_{\rm C2F:25\%}}\sim 0.5\,{\rm pc}$ from the nearest filament (see Fig. \ref{FIG:DistOverRad}a). If these cores have a mean radius of $r_{_{\rm C}}\sim 0.05\,{\rm pc}$ and isothermal sound speed $c_{_{\rm S}}\sim 0.2\,{\rm km\,s^{-1}}$, they will disperse on a timescale $t_{\rm disp}\sim 0.25\,{\rm Myr}$. They must have a bulk velocity $v\!>\!2\,{\rm km\,s^{-1}}$ to reach the filament before they disperse. The lower limit on $v$ is justified on two counts: first, because the actual distance is greater than the projected distance; and second, because the starless core may not be moving towards the nearest point on the filament. Moreover, even if the nearest filament is infinitely long, and if it has a diameter of $0.6\,{\rm pc}$, it only subtends a solid angle $\Omega = \pi(0.6\,{\rm pc}/(0.5\,{\rm pc})$, i.e. $30\%$ of the sky, so if the bulk velocity of the starless core is randomly oriented, it has only a $30\%$ chance of ever reaching the filament.

It is appropriate to consider whether the above estimate should be modified to take account of gravity, which will both accelerate the bulk motion of the core towards the filament, and deflect its trajectory towards the filament (gravitational focussing). The gravitational potential outside a straight filament with line-density $\mu$ is $\phi(r)=2G\mu\ln(r/r_{_{\rm B}})$, where $r$ is the radial distance from the spine of the filament, and $r_{_{\rm B}}$ is the boundary radius of the filament. If, for the purpose of illustration, we adopt (i) a transcritical filament with $\mu =\mu_{_{\rm CRIT}}=2c_{_{\rm S}}^2/G\sim 16\,{\rm M_{_\odot}\,pc^{-1}}$ (where we have substituted $c_{_{\rm S}}=0.2\,{\rm km\,s^{-1}}$, appropriate for molecular gas at temperature $T\sim 10\,{\rm K}$), and (ii) and the canonical $r_{_{\rm B}}\sim 0.3\,{\rm pc}$, then $\phi(r)$ $\sim 4c_{_{\rm S}}^2\ln(r/0.3{\rm pc})$ $\sim 0.16\,{\rm km^2\,s^{-2}}\ln(r/0.3{\rm pc})$. It follows that a starless core which starts off at $r=D_{_{\rm C2F:25\%}}\sim 0.5\,{\rm pc}$ with velocity $v>2\,{\rm km\,s^{-1}}$ directed towards a nearby filament acquires an extra velocity $\Delta v=(G\mu_{_{\rm CRIT}}/v)\ln(D_{_{\rm C2F:25\%}}/r_{_{\rm B}})<0.02\,{\rm km\,s^{-1}}$, which can safely be ignored compared with $v>2\,{\rm km\,s^{-1}}$. Even if we consider the maximum line-density for the filaments in Ophiuchus, $\mu_{_{\rm MAX}}\sim 100\,{\rm M_{_\odot}\,pc^{-1}}$ (see Fig. 5A in \citet{Arzoumanian2019}), the velocity increases by at most $0.2\,{\rm km\,s^{-1}}$, which is still small. Gravitational focussing is not important.

It follows that a core in a filament is effectively trapped by the gravitational potential well of the filament. Its specific thermal energy $3c_{_{\rm S}}^2/2\sim 0.06\,{\rm km^2\,s^{-2}}$ is insufficient to escape from the gravitational potential of the filament. Such cores also have access to a large reservoir of material from which to accrete, and a large external pressure to inhibit their dispersal.

Conversely, many of the starless cores outside of filaments are probably transient condensations that will disperse. Because they are not in filaments, they do not have a large nearby reservoir of material from which to accrete, they do not have a very large external pressure to inhibit their dispersal, and they also do not have the background gravitational field of a filament to inhibit their dispersal.

\begin{table*}
\centering
\caption{The distribution of mass in L1688 and L1689. In each case we give values for the whole sub-region (as defined by the rectangular outlines on Fig. \ref{fig:obsChart}) and for the HERs (which are the regions with surface density exceeding $160\,{\rm M_{_\odot}\,pc^{-2}}$, marked with black contours on Figs. \ref{fig:L1688ContourCore} and \ref{fig:L1689ContourCore}. {\it First row:} the total mass of gas and dust in the region (excluding protostars). {\it Second row:} The total area of the region. {\it Third row:} the mean surface density of the region. {\it Fourth row:} the number of starless and prestellar cores. {\it Fifth row:} The net mass of starless and prestellar cores. {\it Sixth row:} The fraction of mass in starless and prestellar cores.}
\begin{tabular}{lcccccc}\hline
 & $\;\;\;\;$ & \multicolumn{2}{c}{L1688} & $\;\;\;\;$ & \multicolumn{2}{c}{L1689} \\
 && SubR & HER && SubR & HER \\\hline
Total mass && $650\,{\rm M}_{_\odot}$ & $290\,{\rm M}_{_\odot}$ && $400\,{\rm M}_{_\odot}$ & $89\,{\rm M}_{_\odot}$ \\
Total area && $5.7\,{\rm pc}^2$ & $1.2\,{\rm pc}^2$ && $6.0\,{\rm pc}^2$ & $0.37\,{\rm pc}^2$ \\
Mean surface-density && $110\,{\rm M_{_\odot}\,pc}^{-2}$ & $240\,{\rm M_{_\odot}\,pc}^{-2}$ && $66\,{\rm M_{_\odot}\,pc}^{-2}$ & $240\,{\rm M_{_\odot}\,pc}^{-2}$ \\
Number of cores && 288 & 219 && 101 & 81 \\
Net mass of cores && $100\,{\rm M}_{_\odot}$ & $84\,{\rm M}_{_\odot}$ && $41\,{\rm M}_{_\odot}$ & $32\,{\rm M}_{_\odot}$ \\
Fraction of mass in cores && 16\% & 29\% && 10\% & 36\% \\\hline
\end{tabular}
\label{tab:mass}
\end{table*}

\section{Analysis of the Network of Filaments in Rho Oph}\label{sec:filaments}

\subsection{Identifying filaments}

\begin{figure*}
\centering
\includegraphics[width = \textwidth]{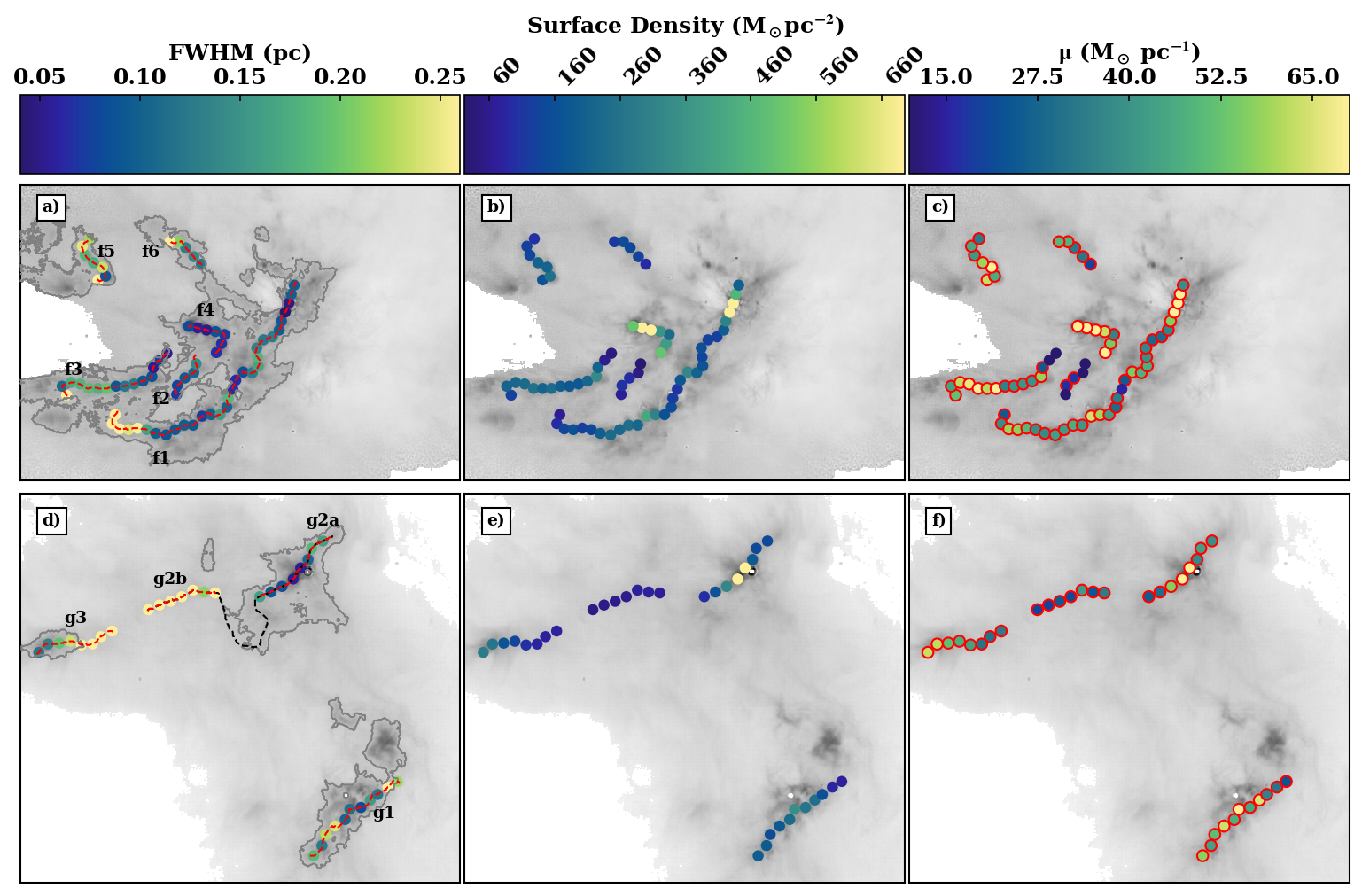}
\caption{Filament parameters for each segment. {\it Top row (a-c):} L1688. {\it Bottom row (d-f):} L1689. {\it Left column (a,d):} colour-coded {\sc fwhm} for each segment. {\it Middle column (b,e):} colour-coded surface density on spine, $\Sigma^0$. {\it Right column (c,f):} colour-coded line-density, $\mu$.}
\label{fig:FilChap}
\end{figure*}

To analyse the networks of filamentary structures visible in the surface density maps, we apply the FilFinder algorithm \citep{Koch2015}, with the minimum threshold (\texttt{glob\_thresh}) set to five times the median background noise (i.e. $\Sigma >180\,{\rm M_\odot\,pc^{-2}}$ for L1688, and $\Sigma >135\,{\rm M_\odot\,pc^{-2}}$ for L1689) and the minimum structure area (\texttt{size\_thresh}) set to 400 pixels (equivalent to \SI{0.009}{\pc \squared}). Next we prune each filament of its branches, leaving the longest path through the filament to trace its spine. Finally we discard filaments shorter than \SI{0.3}{\pc}. These are the selection criteria proposed by \citet{Arzoumanian2011}, with a view to ensuring that all filaments have an aspect ratio $\geq 3$, assuming a width of \SI{0.1}{\pc}. We find six filaments satisfying these criteria in L1688, and three in L1689. The spines of these filaments, along with unique identifiers, are traced on Figs. \ref{fig:FilChap}a and \ref{fig:FilChap}d. FilFinder finds a loop near the middle of the \verb|g2| filament in L1689, which appears to trace the intersection of the filament with the edge of a clump; the loop is shown with a dashed black line on Fig. \ref{fig:FilChap}d. In the sequel we exclude the loop and analyse the two parts on either side of the loop separately, as \verb|g2a| and \verb|g2b|.

We note that in L1688, all six filaments lie within the HERs (shown in Fig. \ref{fig:FilChap}a and \ref{fig:FilChap}d by the grey contours). In L1689, however, only \verb|g1| and \verb|g2a| are completely within HERs. Approximately half of \verb|g3| extends beyond the HER associated with L1689-East, while \verb|g2b| falls in the lower density region between L1689-East and L1689-North.

\subsection{Fitting filament profiles}

To quantify the properties of the filaments, we first define discrete spine points equally spaced at intervals of \SI{7}{\arcsecond} (\SI{\sim 0.005}{\pc}) along the spine of each filament. Next we determine the local tangent to the spine at each spine point by spline fitting. We then compute the column density, at discrete values of the impact parameter, $b$, along the line through the spine point and orthogonal to the local tangent. This gives a local column density profile for each spine point -- albeit a rather noisy one.

Finally we apply the FilChaP\footnote{\url{https://github.com/astrosuri/filchap}} algorithm \citep{Suri2019} to each of the filaments in turn. FilChaP divides a filament up into segments, where each segment is constructed from the average of 12 contiguous local profiles. This greatly improves the signal-to-noise, while retaining the spatial variations in the filament properties that would otherwise be lost by averaging all the local profiles along the entire filament length. The positions of these segments are marked by the coloured circles on Fig. \ref{fig:FilChap}. 

FilChaP performs a background subtraction for each segment, and fits the remaining column-density profile with a Plummer-like profile \citep{Whitworth2001},
\begin{equation}
    \label{EQN:FilChaPPlum}
    \Sigma(b) = \Sigma_0 \left\{1 + \left(\frac{b}{r_0}\right)^2\right\}^{-\frac{p-1}{2}}.
\end{equation}
Here $\Sigma(b)$ is the column density at impact parameter $b$,  $\Sigma_0$ is the column density on the spine, $r_0$ is the radius within which the volume density of the filament is approximately uniform, and $p$ is the radial density exponent outside radius $r_0$. For this study we have fixed $p\!=\!2$ (see \citet{Palmeirim2013}, \citet{Arzoumanian2019} and \citet{Howard2019} for a discussion of this choice, and \citet{Nakamura1999} for a theoretical justification in terms of similarity solutions for the collapse of an approximately isothermal filament). We stress that $\Sigma(b)$, and hence also $\Sigma_0$ and $r_0$, are background-subtracted parameters of the filaments. FilChaP estimates the local line-density of the filament, $\mu$, by summing explicitly the contributions to the background-subtracted filament profile (not by integrating the fitted profile, Eq. \ref{EQN:FilChaPPlum}).

Once FilChaP has returned estimates of $\Sigma_0$, $r_0$ and $\mu$, for each segment along a filament, we compute the local \textsc{fwhm}$\,=2(2^{1/2}-1)^{1/2}r_0=1.287\,r_0$, and this is plotted on Figs. \ref{fig:FilChap}a (L1688) and \ref{fig:FilChap}d (L1689). The median \textsc{fwhm} of all the filaments in L1688 is \SI[separate-uncertainty=true]{0.11 \pm 0.04}{\pc}, while the median \textsc{fwhm} of the L1689 filaments is \SI[separate-uncertainty=true]{0.19 \pm 0.08}{\pc}. However, the filament segments that fall outside the HERs in L1689 are significantly and systematically wider than those sections within the HERs. If we exclude these low density segments, the median filament \textsc{fwhm} in L1689 decreases to \SI[separate-uncertainty]{0.14 \pm 0.05}{\pc}. These median \textsc{fwhm}s are larger than those found by \citet{Howard2019} for the Taurus L1495 filament. The median \textsc{fwhm}s of the filaments are given in Table \ref{tab:filaments}, and span a wide range from \SI[separate-uncertainty=true]{0.066 \pm 0.002}{\pc} for \verb|f4| in L1688, to \SI[separate-uncertainty=true]{0.315 \pm 0.008}{\pc} for \verb|g2b| in L1689. We note that the {\sc fwhm}s we find are significantly larger than those estimated by \citet{Arzoumanian2019}, who find a mean and standard deviation of $0.06\pm0.02\,{\rm pc}$ in Ophiuchus; the reason for this difference is unclear.

\begin{table}
\centering
\caption{The properties of each of the filaments identified in the L1688 and L1689 sub-regions. $\Sigma_0$, \textsc{fwhm} and $\mu$ are length averaged median values.}
\begin{tabular}{c|ccccc}\hline
\multicolumn{5}{c}{L1688} \\\hline
Fil & $\Sigma_0$ & \textsc{fwhm} & ${\mu}$  & Length \\
& $\left({\rm M_\odot\,pc^{-2}}\right)$ & $\left(0.1\,{\rm pc}\right)$ & $\left({\rm M_\odot\,pc^{-1}}\right)$ & $\left(\rm{pc}\right)$ \\\hline
f1 &185$\pm$8  &1.00$\pm$0.04 &42$\pm$2 & 1.76\\
f2 &70$\pm$4  &0.95$\pm$0.02 &10.5$\pm$1.4 & 0.28\\
f3 &215$\pm$4  &1.21$\pm$0.08 &43.6$\pm$4.5 & 0.81\\
f4 &506$\pm$21 &0.66$\pm$0.02 &70.1$\pm$5.9 & 0.35\\
f5 &171$\pm$6  &2.21$\pm$0.06 &46.0$\pm$4.7 & 0.35\\
f6 &139$\pm$3   &1.39$\pm$0.03 &32.8$\pm$5.7 & 0.24\\\hline
\multicolumn{5}{c}{L1689} \\\hline
Fil & $\Sigma_0$ & \textsc{fwhm} & ${\mu}$  & Length \\
& $\left({\rm M_\odot\,pc^{-2}}\right)$ & $\left(0.1\,{\rm pc}\right)$ & $\left({\rm M_\odot\,pc^{-1}}\right)$ & $\left(\rm{pc}\right)$ \\\hline
g1  &195$\pm$7 &1.58$\pm$0.08 &46.9$\pm$3.5 & 0.61\\
g2a &179$\pm$8 &0.98$\pm$0.07 &41.3$\pm$4.4 & 0.46\\
g2b &62$\pm$2  &3.15$\pm$0.08 &23.8$\pm$1.3 & 0.36\\
g3  &122$\pm$7 &2.60$\pm$0.03 &45.4$\pm$4.4 & 0.42\\\hline
\end{tabular}
\label{tab:filaments}
\end{table}

The values of $\Sigma_0$ for each segment are shown in Fig. \ref{fig:FilChap}b for L1688, and Fig. \ref{fig:FilChap}e for L1689. Filament \verb|f4| clearly has a higher average central column density than any of the other filaments in either sub-region. Filament \verb|f1| in L1688 and filament \verb|g2a| in L1689 both contain a small number of contiguous segments which exhibit a much higher central column density than the rest of the filament. For \verb|f1|, these segments are located near the line connecting the stars S1 and HD147889, indicating that local feedback from these B stars may be significantly squeezing the filament at this location, enhancing its column density; this possibility is discussed further in Section \ref{sec:f1feedback} below. The more dense segments in \verb|g2a| are located near 1689-IRS6, which may indicate that local feedback is also a factor in the density increase there. Filament \verb|g2b| and the segments of \verb|g3| that lie outside the HERs show a decrease in central density which mirrors the increase in filament width discussed previously. The median central column density of each of the filaments is listed in Table \ref{tab:filaments}.

\subsection{Filament line-densities}

Figs. \ref{fig:FilChap}c and \ref{fig:FilChap}f show the line-densities, $\mu$, for the filament segments in L1688 and L1689, respectively. The median segment line-density in L1688 is \SI{47}{\msol \per \pc}, with a 32/68-percentile range of \SI{\pm 18}{\msol \per \pc}. The median segment line-density in L1689 is \SI{46}{\msol \per \pc}, with a 32/68-percentile range of \SI{\pm 15}{\msol \per \pc}. The 32/68-percentile range is equivalent to the standard deviation for a Gaussian distribution, and thus acts as the standard deviation for a median value. The median line densities for each filament are listed in Table \ref{tab:filaments}. 

If we assume that the gas is isothermal with isothermal sound speed $c_{_{\rm S}} = \,\SI{0.19}{\kilo \meter \per \second}$ (corresponding to molecular gas at \SI{11}{\kelvin}), the critical line-density \citep{Ostriker1964, Inutsuka1997} is 
\begin{equation}
    \label{EQN:mucrit}
    \mu_{_{\rm C}} = \frac{2 c_{_{\rm S}}^2}{G} \simeq 16.2\,\si{\msol \per \pc}.
\end{equation}
In the absence of other support mechanisms (e.g., magnetic fields or turbulence), isothermal filaments with $\mu > \mu_{_{\rm C}}$ should collapse and fragment under gravity. In Figs. \ref{fig:FilChap}c and \ref{fig:FilChap}f, we outline segments with $\mu > \mu_{_{\rm C}}$ in red. This implies that all the filaments in both sub-regions are super-critical, and -- in the absence of other support mechanisms -- highly prone to collapse and fragmentation. However, the Green Bank Ammonia Survey finds that the median gas-kinetic temperature of dense cores in L1688 is $17.5(\pm 5.3)\,{\rm K}$ \citep{Kerr2019}, and we might expect that the somewhat less dense gas in the filaments to be even warmer. The dust temperatures returned by PPMAP are also significantly higher than \SI{10}{\kelvin}, but the densities in the filaments analysed here are probably too low for close thermal coupling between gas and dust. If the gas temperature is \SI{25}{\kelvin} ($c_{_{\rm S}} = \,\SI{0.30}{\kilo \meter \per \second}$), then $\mu_{_{\rm C}} =\, \SI{40.5}{\msol \per \pc}$, and just 56\% of the segments in L1688, and 51\% of the segments in L1689, are super-critical.

We note that \citet{Arzoumanian2019} do not correct the line densities of the filaments they analyse for the contribution from embedded cores, on the grounds that this contribution is small. It is worth noting that, even if this contribution is not small, it should not be discounted. Hydrostatic equilibrium only exists for an isothermal filament if the outward acceleration due to internal pressure, $\sim 4c_{_{\rm S}}^2/r$ balances the inward acceleration due to gravity, $\sim 2G\mu/r$. Any inward gravitational acceleration, be it due to relatively diffuse gas, dense cores or even protostars and stars, must be balanced for there to be an equilibrium. So $\mu$ should represent contributions from all sources of gravity inside the filament.

\subsection{The Influence of S1 and HD147889 on Filament f1}\label{sec:f1feedback}
As mentioned previously, the B stars S1 and HD147889 appear to influence the dense material in L1688. In particular, the section of \verb|f1| between the two stars has a much narrower width, and a much higher column density, than the rest of the filament, and coincides with a concentration of cores.

\begin{figure}
    \centering
    \includegraphics[width = \linewidth]{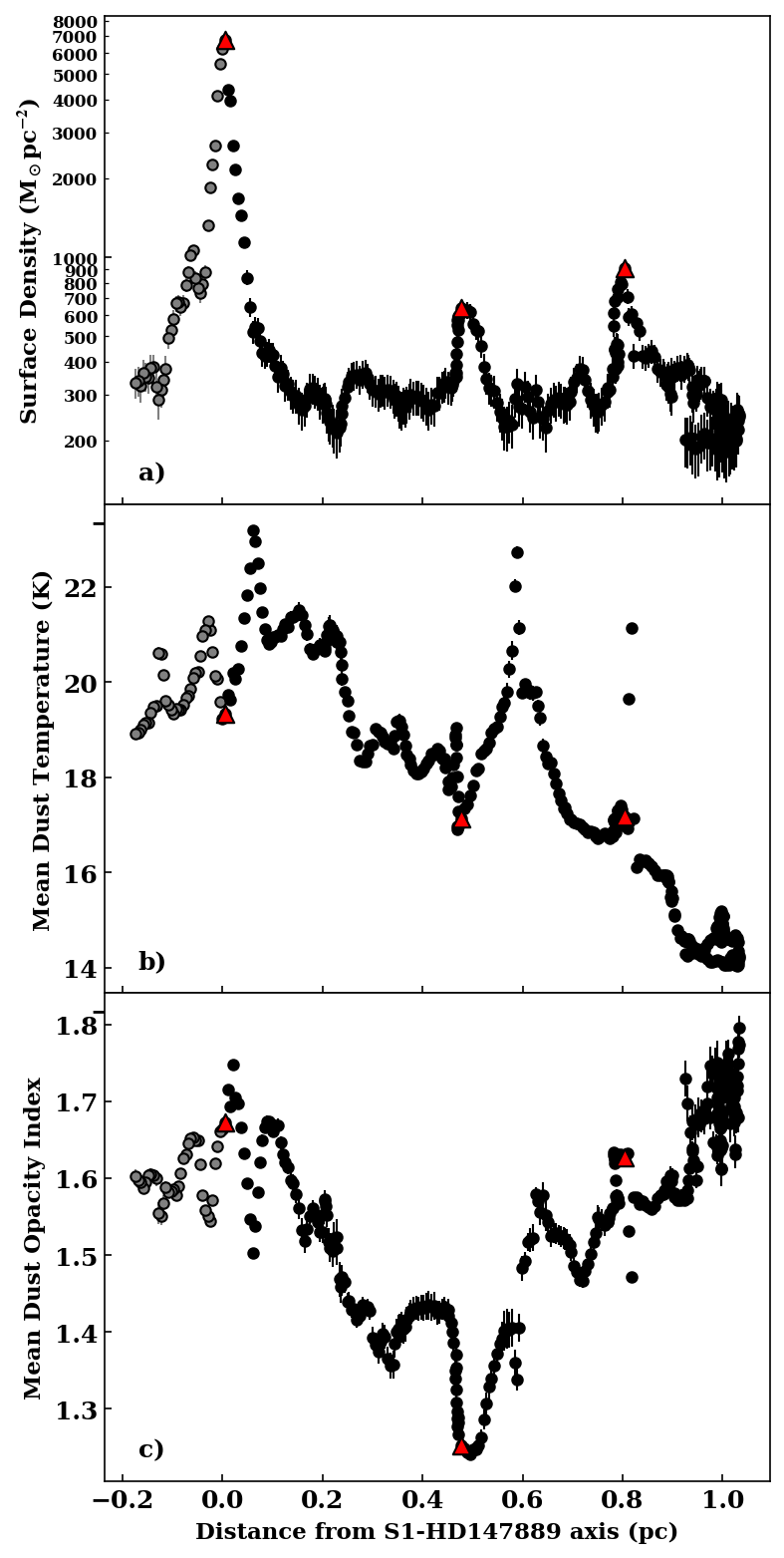}
    \caption{Plots of (a) surface density, $\Sigma_0$, (b) mean dust temperature, $\bar{T}$, and (c) mean dust opacity index, $\bar{\beta}$, for the spine points of filament \texttt{f1} in L1688, plotted against projected distance from the \AoI $\;$(see red line on Fig. \ref{fig:L1688ContourCore}). Grey-filled circles at negative value of the abscissa represent points north-west of this axis, and black-filled circles at positive value of the abscissa represent points to the south-east. Filled red triangles indicate the main local column-density maxima discussed in the text.}
    \label{fig:f1feedback}
\end{figure}

To quantify this, we define the straight line connecting S1 and HD147889 (the red line on Fig. \ref{fig:L1688ContourCore}) as the \AoI. On Fig. \ref{fig:f1feedback} we plot the surface density, $\Sigma_0$, the mean dust temperature, $\bar{T}$, and the mean dust opacity index, $\bar{\beta}$, at each spine point, against its distance from the \AoI. The black filled circles represent spine points that lie on the south-east side of the \AoI, and the grey filled circles represent those that lie on the other side. The red outlined points correspond to local maxima in the column density. 

Fig. \ref{fig:f1feedback}a shows a strong peak in the surface density for spine points close to the \AoI, with a maximum of $\sim 680\,{\rm M_\odot\,pc^{-2}}$ actually on the axis. $\Sigma_0$ then decreases rapidly with distance from the \AoI, dropping by more than an rder of magnitude over \SI{0.2}{\pc} on either side. Beyond this distance, $\Sigma_0$ is approximately constant at $300\pm 100\,{\rm M_\odot\,pc^{-2}}$. There are well-defined local peaks at \SI{\sim0.5}{\pc} and \SI{\sim 0.8}{\pc} from the \AoI, in both cases corresponding to features on the south-east side.

The column-density peak on the \AoI$\,$ presumably reflects enhanced pressure from the regions surrounding S1 and/or HD147889 acting to compress the material within the filament. As shown on Fig. \ref{fig:L1688ContourCore}, this region has an exceptionally high density of cores, both near to, and along the filament spine. However, the width of the filament here is quite low, and -- if one discounts the mass in cores -- the line-density is actually less supercritical than many other filament sections in Ophiuchus. This suggests that compression by feedback from the nearby B stars may have induced efficient fragmentation. This scenario is further supported by the morphological features visible in the column-density and mean-temperature maps (Figs. 4 and 5a), which suggest compression waves emanating from the two B stars, especially S1.

Fig. \ref{fig:f1feedback}b shows that the mean dust temperature on the spine points tends to fall with distance from the \AoI, implying that one or other of the PMS stars, and possibly both, are having a significant heating effect on the filaments in their vicinity. Local peaks and troughs in the mean temperature presumably reflect locations where the dust in the filament is -- respectively -- exposed to additional radiation (for example, due to an embedded protostar) or more strongly shielded (for example, due to a dense but starless core). For example, at the spine points close to the \AoI, and at the spine points \SI{\sim 0.5}{\pc} from the \AoI, column-density maxima coincide with minima in the mean dust temperature, but are closely flanked by maxima in the mean dust temperature, suggesting dense clumps or cores that have cold interiors but strongly externally heated envelopes.

There is no systematic correlation between the mean dust opacity index and distance from the \AoI $\;$on Fig. \ref{fig:f1feedback}c, suggesting that the increased density and dust temperature in this region has not yet had any discernible influence on the mean physical or chemical properties of the dust. There is, however, a  clear decrease in the dust opacity index at \SI{\sim 0.5}{\pc}, coincident with the local column-density maximum there.

\section{Conclusions}\label{sec:conc}

We have analysed \textit{Herschel} and SCUBA-2 observations of the L1688 and L1689 sub-regions of the Ophiuchus molecular cloud, using the PPMAP algorithm. PPMAP returns a 4D data-hypercube giving, for each pixel on the sky, the surface density, $\Sigma$, as estimated from the opacity of dust of different types (represented by different opacity indices, $\beta$) and at different temperatures ($T$). Throughout both sub-regions, the PPMAP results show a network of filaments with complex temperature-coherent structures, and a systematic change in the dust properties in high-density regions, similar to that observed in the Taurus L1495 filament by \citet{Howard2019}. We note the following features.
\begin{enumerate}
    \item The L1688 and L1689 sub-regions (defined, somewhat arbitrarily, by the black rectangles on Fig. \ref{fig:obsChart}) have average surface densities of \SI{\sim 110}{\msol \per \pc \squared} and \SI{\sim 66}{\msol \per \pc \squared} respectively.
    \item $\sim 44\%$ of the mass in the L1688 sub-region is contained within High Efficiency Regions (HERs), defined as regions with $\Sigma\geq 160\,{\rm M_\odot\,pc^{-2}}$ (equivalent to $N_{_{\rm H_2}}\geq 7 \times 10^{21}\,{\rm cm}^{-2},\;\si{\Av} \geq 7$). HERs make up $\sim 22\%$ of the mass in the L1689 sub-region.
    \item Taking the two sub-regions together, most starless and pre-stellar cores (300 out of 389) lie within the HERs, as already noted by \citet{Ladjelate2020}. 
    \item Starless and prestellar cores account for 29\% of the mass in the L1688 HERs, and 36\% in the L1689 HERs.
    \item These results confirm the conclusion of \citet{Konyves2015} that there is a column-density threshold for efficient core formation, $\Sigma\geq 160\,{\rm M_\odot\,pc^{-2}}$ (equivalent to $N_{_{\rm H_2}}\geq 7 \times 10^{21}\,{\rm cm}^{-2},\;\si{\Av} \geq 7$); and they suggest that, once this threshold is exceeded, the core formation efficiency is approximately universal at $\eta\!\sim\!0.5$, regardless of the wider environmental properties.
    \item The filaments identified in L1688 all lie completely within HERs, while in L1689, two filaments fall entirely within HERs, one filament is partially located within an HER, and the final filament is found in the low density region bridging two HERs. 
    \item The filaments in L1688 are narrower, $\textsc{fwhm}$ $=$ $0.11(\pm 0.04)\,{\rm pc}$, than those in L1689, $\textsc{fwhm}$ $=$ $0.19(\pm 0.08)\,{\rm pc}$. However, after discounting the sections of filaments in L1689 that lie outside the HERs, the average filament width in L1689, $\textsc{fwhm}$ $=$ $0.14(\pm 0.05)\,{\rm pc}$, is closer to those in L1688.
    \item The mean line-density of the filaments in the two sub-regions are similar: $47(\pm18)\,{\rm M_{_\odot}\,pc^{-1}}$ for L1688 and $46(\pm15)\,{\rm M_{_\odot}\,pc^{-1}}$ for L1689. If we adopt the canonical critical line density, \SI{16.2}{\msol \per \pc}, which presumes molecular gas at \SI{10}{\kelvin} and ignores other forms of support (magnetic fields and/or turbulence), the filaments in both sub-regions are highly super-critical and should fragment rapidly.
    \item On the basis of the Green Bank Ammonia Survey, \citet{Kerr2019} estimate that the dense cores in L1688 have a mean gas-kinetic temperature of \SI{\sim 18}{\kelvin}, and the somewhat less dense material in the filaments may be even warmer. If the gas-kinetic temperature in the filaments is \SI{25}{\kelvin}, the critical line-density is \SI{40.5}{\msol \per \pc}. The filaments are then trans-critical, and fragmentation should be patchy. (Although the dust temperatures in the filaments are typically in the range $\sim 17\,{\rm K}$ to $\sim 24\,{\rm K}$, the densities are probably too low for the gas and dust to be well coupled thermally.)
    \item Feedback from the B stars S1 and HD147889 appears to be compressing and heating a section of the \texttt{f1} filament in L1688, where it passes between them, accelerating the rate of core formation there.
\end{enumerate}

\section*{Acknowledgements}

We thank the referee, Doris Arzoumanian, for her very thorough and thoughtful report, which helped us to improve the original version of this paper. This research made use of Photutils, an Astropy package for detection and photometry of astronomical sources \citep{Bradley2016}. ADPH gratefully acknowledges the support of a PhD studentship from the UK Science and Technology Facilities Council (STFC). APW, MJG and MWLS gratefully acknowledge the support of a Consolidated Grant (ST/K00926/1) from the STFC. 

\section*{Data Availability}
The observational data underpinning this article are collected from the \Herschel{} Space Observatory and the James Clerk Maxwell Telescope (JCMT), and are publicly available from \Herschel{} Science Archive (\url{http://archives.esac.esa.int/hsa/whsa}) and the JCMT Gould Belt Survey (\url{https://doi.org/10.11570/18.0005}), respectively. The core catalogue presented in \citet{Ladjelate2020} is publicly available from \url{doi.org/10.26093/cds/vizier.36380074}. The derived data generated in this research will be shared on request to the corresponding author.

\bibliographystyle{mnras}
\bibliography{OphiuchusPPMAP.bib}

\appendix

\section{Point Process Statistical Degeneracy}\label{apx:PPSD}

As discussed in Sec. \ref{sec:ppmap-prod}, the PPSD (Eq. \ref{EQN:PPSD}) gives a pixel-by-pixel measure of the statistical significance of the PPMAP optical-depth estimates, and hence also of the surface-density estimates. It is analogous to a signal-to-noise ratio, with $\mathrm{PPSD} = 1$ indicating that the magnitude of an optical-depth estimate is equal to the magnitude of the uncertainty associated with that estimate, whereas $\mathrm{PPSD} = 5$ indicates that the estimate is five times larger than the associated uncertainty, and so on.

\begin{figure*}
    \centering
    \includegraphics[width = 0.95\textwidth]{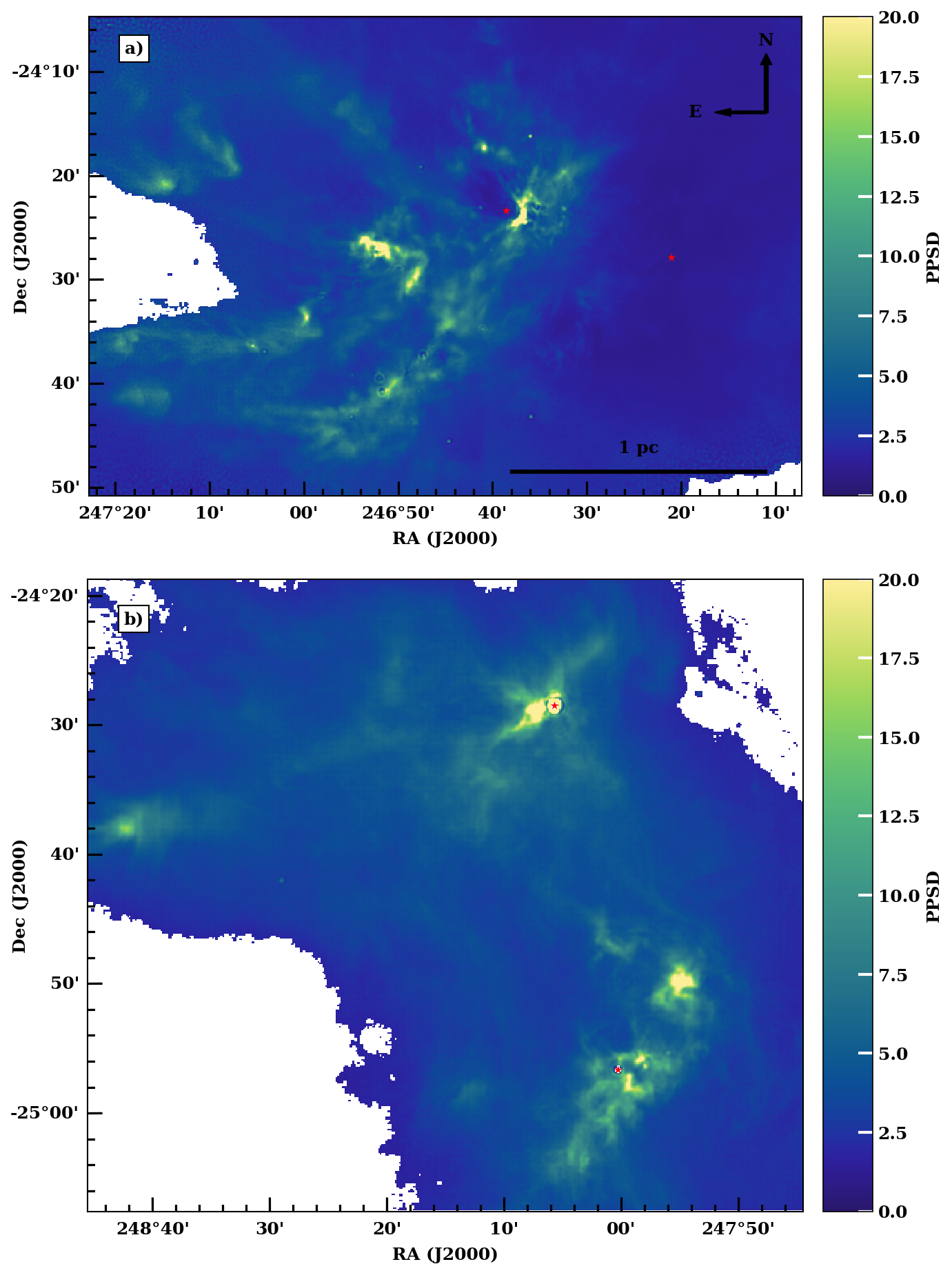}
    \caption{Pixel-by-pixel PPSD for (a) the L1688 sub-region, (b) the L1689 sub-region.}
    \label{fig:PPSD}
\end{figure*}

Figs. \ref{fig:PPSD}a and \ref{fig:PPSD}b show the PPSD values for the L1688 and L1689 sub-regions of Ophiuchus. The mean PPSD value is 2.9 for the L1688 sub-region, and 3.9 for the L1689 sub-region. The PPSD for the L1688 HER is 5.3, and 8.1 for the L1689 HER.

\bsp	
\label{lastpage}
\end{document}